\documentclass[a4paper,11pt]{article}
\usepackage[dvipsnames]{xcolor}
\usepackage[utf8]{inputenc}
\usepackage[bbgreekl]{mathbbol}
\usepackage{geometry}
\usepackage{makecell}

\usepackage{xcolor}
\usepackage{amsmath, array, amssymb, amsfonts,amsthm}
\usepackage{upgreek}
\usepackage{xfrac}
\usepackage[inline]{enumitem}
\setlist{nolistsep}
\usepackage[french, english]{babel}
\usepackage[all]{xy}
\usepackage{txfonts}  
\usepackage{sectsty}
\usepackage{booktabs}
\usepackage{caption}
\usepackage{dsfont}
\usepackage{mathtools}
\usepackage{slashed}
\usepackage[makeroom]{cancel}
\usepackage[hidelinks]{hyperref}
\usepackage{textcomp}
\usepackage{multicol}
\usepackage[title]{appendix}
\usepackage[square,numbers,sort,compress,semicolon,merge]{natbib}
\let\cite\citep 
\usepackage{hyperref}
\hypersetup{colorlinks=true, urlcolor=blue, citecolor=blue, linktoc=page}

\usepackage{tikz}
\usepackage{tikz-cd}
\usetikzlibrary{cd}
\tikzcdset{
arrow style=tikz,
diagrams={>={Straight Barb[scale=0.8]}}
}

\DeclareMathAlphabet{\mathpzc}{OT1}{pzc}{m}{it}

\usepackage{mathrsfs}
\usepackage{footmisc}

\usepackage{mathtools}

\makeatletter
\renewcommand*\env@matrix[1][\arraystretch]{%
  \edef\arraystretch{#1}%
  \hskip -\arraycolsep
  \let\@ifnextchar\new@ifnextchar
  \array{*\c@MaxMatrixCols c}}
\makeatother

\newcommand{\defeq}{\vcentcolon=}

\renewcommand\P{\mathcal{P}}

\newcommand\R{\mathcal{R}}
\newcommand\RR{\mathbb{R}}
\newcommand\CC{\mathbb{C}}

\renewcommand\1{\textbf{1}}

\newcommand\T{\mathcal{T}}
\newcommand\G{\mathcal{G}}
\renewcommand\H{\mathcal{H}}
\newcommand\A{\mathcal{A}}

\renewcommand\L{\mathcal{L}}
\renewcommand\S{\mathcal{S}}

\newcommand\SO{\mathcal{SO}}

\newcommand\vphi{\varphi}

\newcommand\sS{\mathsf{S}}

\renewcommand\epsilon{\varepsilon}

\newcommand\rarrow{\rightarrow}
\newcommand\xrarrow{\xrightarrow}

\newcommand\aut{\mathfrak{aut}}

\newcommand\LieG{\mathfrak{g}}
\newcommand\LieH{\mathfrak{h}}

\newcommand\su{\mathfrak{su}}
\newcommand\so{\mathfrak{so}}

\newcommand\co{\mathfrak{co}}
\newcommand\gl{\mathfrak{gl}}

\renewcommand\t{\tilde}

\renewcommand\b{\bar }
\newcommand\w{\wedge}
\renewcommand\d{\partial}
\newcommand\s{\sigma}
\newcommand\bs{\boldsymbol}

\renewcommand\-{^{-1}}
\newcommand\Ad{\text{Ad}}
\newcommand\ad{\text{ad}}

\renewcommand\1{\mathds{1}}

\makeatletter

\newcommand{\Rmnum}[1]{\expandafter\@slowromancap\romannumeral #1@}
\makeatother

\makeatletter
\newcommand{\leqnomode}{\tagsleft@true\let\veqno\@@leqno}
\newcommand{\reqnomode}{\tagsleft@false\let\veqno\@@eqno}
\makeatother

\newcommand\diff{\mathfrak{diff}}

\DeclareMathOperator{\Diff}{Diff}
\DeclareMathOperator{\Aut}{Aut}
\DeclareMathOperator{\Tr}{Tr}

\theoremstyle{definition}

\makeatother

\begin{document}

\title{Cartan geometry, supergravity \\ and group manifold approach}
\author{J. François${\,}^{a,\,b,\,c}$, L. Ravera${\,}^{d,\,e,\,f}$}
\date{\today}

\maketitle
\begin{center}
\vskip -0.8cm
\noindent
${}^a$ Department of Mathematics \& Statistics, Masaryk University -- MUNI. \\
Kotlářská 267/2, Veveří, Brno, Czech Republic. \\[2mm]
${}^b$ Department of Philosophy -- University of Graz. \\
Heinrichstraße 26/5, 8010 Graz, Austria. \\[2mm]
${}^c$ Department of Physics, Mons University -- UMONS.\\
Service \emph{Physics of the Universe, Fields \& Gravitation}. \\
20 Place du Parc, 7000 Mons, Belgium. \\[2mm]
${}^d$ DISAT, Politecnico di Torino -- PoliTo. \\
Corso Duca degli Abruzzi 24, 10129 Torino, Italy. \\[2mm]
${}^e$ Istituto Nazionale di Fisica Nucleare, Section of Torino -- INFN. \\
Via P. Giuria 1, 10125 Torino, Italy. \\[2mm]
${}^f$ \emph{Grupo de Investigación en Física Teórica} -- GIFT. \\
Concepción, Chile. 
\end{center}

\vspace{-3mm}

\begin{abstract}
We make a case for the unique relevance of Cartan geometry for gauge theories of gravity and supergravity. 
We introduce our discussion by recapitulating historical threads, providing motivations. 
In a first part we review the geometry of classical gauge theory, as a background for understanding gauge theories of gravity in terms of Cartan geometry. 
The second part introduces the basics of the group manifold approach to supergravity, hinting at the deep rooted connections to Cartan supergeometry. 
The contribution is intended, not as an extensive review, but as a conceptual overview, and hopefully a bridge between communities  in  physics and mathematics. 
\end{abstract}


\vspace{-3mm}

\tableofcontents

\bigskip


\section{Introduction}



It is by now well known that the differential geometry of connections on fiber bundles is the mathematical underpinning of classical gauge field theory, which is itself the framework nesting the best current theories of fundamental physics accounting for the structure and interactions of the elementary constituents of the universe. 

Gauge field theories of the Yang-Mills type grew out of General Relativity (GR), largely via the key contributions of H. Weyl between 1918-19 \cite{Weyl1918, Weyl1919} and 1929 \cite{Weyl1929}, in the Abelian $U(1)$ case. In 1954, Yang and Mills gave the first non-Abelian model, with $S\!U(2)$, hence the name synonymous with non-gravitational gauge theories: Yang-Mills (YM) theories \cite{Yang-Mills1954}. A fact quite unfair, as it is the Japanese physicist Ryoyu Utiyama who, almost simultaneously and independently of  Yang and Mills,  laid out the general framework of gauge field theory (including the case of gravity) for any Lie group \cite{Utiyama1956}. See \cite{ORaif1997} for an enlightening concise history of the subject.  

The development of the theory of connections and of fiber bundles also owes much to the impulse given by GR \cite{CartanEspGen24}. It reached full maturity in the 50s \cite{Ehresmann47, Ehresmann1950, Steenrod1951, Kobayashi1957, Marle}, and textbook material in the 60s \cite{Kob-NomI, Kob-NomII, Kobayashi1972}. By the late 70s, it had become clear that it is the rigorous foundation of classical gauge field theory \cite{Wu-Yang1975, Yang1976, Eguchi-et-al980, Singer1978, Singer1981}. 
This was slowly assimilated by the physics community, bore fruitful results in the 80s in the realm of non-perturbative field theory \cite{Cotta-Ramusino-Reina1984} (e.g. regarding gauge anomalies \cite{Bonora-Cotta-Ramusino, Stora1984, Catenacci-et-al1986, Catenacci-Pirola1990, Bertlmann, Fines, Bonora2023}), and was certainly common knowledge by the early 90s. We remind the elementary facts of the matter in section 
\ref{Yang-Mills theories: Kinematics and dynamics}.
\medskip

 The origin of classical gauge theories of gravity could be traced back to Einstein's 1925 tetrad formulation introducing local Lorentz, $S\!O(1,3)$, symmetry transformations \cite{Unzicker-Case2005, Sauer2006}. It partly inspired Weyl's 1929 paper \cite{Weyl1929}, dealing with $S\!L(2, \CC)$ spinors, which also cemented in history the gauge principle for $U(1)$ and electromagnetism. The subject started in earnest, in a modern form,  with the 1955 paper of R. Utiyama \cite{Utiyama1956}. It continues with the famous papers by  Kibble 1961 \cite{Kibble1961} and  Sciama 1964 paper \cite{Sciama1964}, reintroducing torsion alongside Riemann curvature in a Lorentz-gauge reformulation of GR. Other variants of GR followed through the 70s, like Poincaré gravity, affine gravity, Weyl-Cartan gravity, de Sitter and conformal gravity -- see ee.g. \cite{Hehl-Kibble-Blagojevic2013} for an introductory overview.  

Gauge theories of gravity do not fit so well in the mold of Ehresmann connections on principal bundles. Their right mathematical home is to be found with Cartan's notion of connection, established between 1923-1925 \cite{CartanConf23, CartanProj24, CartanEspGen24}, as reformulated in the bundle language by Ehresmann \cite{Ehresmann1950},\footnote{Cartan was the teacher of Ehresmann. It is  as he was giving a modernised definition of his master's notion of connection that he proposed his own far reaching generalisation . See  \cite{Marle} for a brief history.} 
and then by Kobayashi in \cite{Kobayashi1955, Kobayashi1957} (around the same time Utiyama was founding the topic of gauge gravity). 
Subsequently, Cartan geometry, though a well-developed subject \cite{tanaka_1957, Kobayashi1961, Kobayashi-Nagano1964, Ogiue, Ochiai, Kobayashi1972}, was relatively eclipsed by the far reaching  development of the general theory of principal connections on bundles and did not reach the same level of widespread awareness. Some nonetheless noticed early their relevance to Penrose's then new twistor theory \cite{Penrose-McCallum72, Penrose1977, Friedrich77}, as well as to non-relativistic classical physics \cite{Duval-et-al1982, Duval-et-al1983}.

In the late 80s and 90s,  the schools  developing  tractor calculi for the classical projective and conformal, and many other, geometries (see e.g. \cite{Bailey-Eastwood91, Bailey-et-al94, Gover99}), 
and the theory of invariant operations for parabolic geometries (e.g. \cite{Cap-et-al1997-1, Cap-et-al1997, Cap-et-al2000}),
were well aware of their deep relations to Cartan connections \cite{Alekseevsky-Michor1994}.  Cartan geometry was first brought to a wider audience by  Sharpe's '97 textbook \cite{Sharpe}, and a more advanced and complete presentation  -- stressing in particulat the organic  relation to tractor calculi for parabolic geometries  -- is the 2009 volume \cite{Cap-Slovak09} by Čap \& Slovák. 

 During that same period, 
 the deep relevance of Cartan connections to gravitational physics was only noticed in isolated instances -- see e.g. \cite{Zardecki1988, Korz-Lewand-2003, Wise07, Wise09, Wise10, Lazzarini2008, AFL2016_I, Attard-Francois2016_II, Francois2021, Francois2021_II}. 
 It is our intention to make the case, in section \ref{Classical gauge theory of gravity and Cartan geometry}, that Cartan geometry is indeed the natural mathematical foundation of classical gauge theories of gravity, so that it should be more widely recognized as such.
\medskip

In quantum field theory (QFT), fermions fields are to be treated as anticommuting variables. Berezin pioneered in the late 60s and 70s the explicit use of Grassmann algebras \cite{Berezin1966}, and super-Lie groups with Kac \cite{Berezin-Kac1970}, to give a firmer mathematical ground to the quantization procedure for theories with fermionic fields, and to try and make sense of the connection between their spin and anticommutation properties \cite{Berezin-Marinov1977}. Furthermore, the Faddeev-Popov \cite{Faddeev-Popov-67} and Becchi-Rouet-Stora-Tyutin (BRST) \cite{BRS-75, BRS-76, Tyutin-75} quantization procedure involves non-physical anticommuting variables called \emph{ghost fields}. 

Jumping to more tentative theoretic constructions, the 70s  saw the birth of the idea of supersymmetric field theories  \cite{Golfand-Likhtman1971, Volkov-Akulov1973, Wess-Zumino1974, Salam-Strathdee1978}. This class of theories have several desirable properties (notably regarding their quantization, with tame divergences), allowing in particular to circumvent the so-called Coleman-Mandula theorem \cite{Coleman-Maudula1967} 
%
%
stating that the (non-Grassmannian) Lie group of symmetries  of a quantum field theory can only be a direct product of spacetime (Poincaré) and internal (Yang-Mills) symmetries. 

It follows that theories with \emph{local} (``gauged" in the physics terminology) supersymmetry  
 naturally are gauge theories of gravity, called supergravity (SUGRA) theories. 
In the late 70s and early 80s, supergravity models  \cite{Cham-West1977, McDowell-Mansouri1977, Ferrara-et-al1977, Kaku_et_al1977, Kaku-et-al1978, Stelle-West1979} became a hot topic, and their geometric interpretation in terms of ``superspaces" was quick to emerge \cite{Wess-Zumino1977}. In particular the ``(super) group-manifold approach" to supergravity was pioneered by Neeman and Regge \cite{Neeman-Regge1978, Neeman-Regge1978b} and developed by the school of Castellani, D'Auria, and  Fré \cite{Castellani-et-al1991}. 
Supersymmetry and supergravity caught additional intellectual wind after their close connection to super-String and M-theory was established between the mid 80s and the mid 90s.

Meanwhile, mathematicians thus took early interest in the late 70s in developing ``supergeometry", i.e. a $\mathbb Z_2$-graded generalisation of standard differential geometry.  As a case of ``mildly" non-commutative geometry, a natural language to do so was via the tools of algebraic geometry, sheaf theory, and locally ringed spaces. This approach is associated to the names of pioneers like Berezin \cite{Berezin1987}, Leites \cite{Leites1980}, Kostant \cite{Kostant1977}). 
An (essentially) equivalent  approach was developed, pioneered by deWitt and Rogers \cite{DeWitt1984,Rogers1980, Yates1980, Rogers2007}, which parallels the standard treatment of differential geometry, defining supermanifold as being locally isomorphic to a model supervector space $\RR^{p|q}$. It is closer to the practice in ``superspace" of physicists. 
Few pursued an even more abstract categorical approach \cite{Shvarts1984, Voronov1984, Molotkov1984-2010}, that would prove able to encompass both previous approaches \cite{Schmitt1997, Sachse2008} and is arguably more apt to handle infinite dimensional superspaces \cite{Sachse-Wockel2012} (the locally ringed space approach is usually deemed maladapted for the task \cite{Deligne-et-al-1999}) -- which are important when studying automorphism and symmetry groups of (super) geometric structures.\footnote{One may draw an analogy from physics,  with the development of quantum mechanics (QM): Its first incarnation as ``matrix mechanics" devised in 1924-25 by Heisenberg, Born and Jordan used mathematics unfamiliar to physicists (matrices). The second version devised by Schrödinger in 1925-26 using standard differential equation was welcomed as more familiar and intuitive. Dirac is credited for first convincingly argue that both approaches were equivalent, but it took Von Neumann to give the broad encompassing (and now textbook) formulation in terms of operators on (possibly infinite dimensional) Hilbert spaces. 
Later developments in terms of $C^*$-algebras are  considered more advanced formulation of QM, and the basis of  refined mathematical formulation of quantum field theory (QFT); one may wonder what would be the analogue in the development of (higher) supergeometry.}  
 \medskip

In the same way that Cartan geometry is the foundation of classical gauge theories of gravity, one expects that Cartan supergeometry is the right mathematical framework for supergravity theories.\footnote{At least the class for which the gravitationnal superpotential is a still a 1-forms: As in string theory, the broader class includes higher form fields necessitating the framework of higher geometry. We come back to this point in the conclusion.} 
Despite the idea being fairly natural -- one of us entertaining it for several years --  in the physics literature, awareness of the relevance of Cartan's framework to supergravity has been  rare and sporadic   \cite{Zardecki1993, Egeileh-ElChami2012}. As a matter of  fact, Cartan supergeometry was never developed in earnest by mathematicians.\footnote{Except in a programmatic way within the considerable \cite{Schreiber2014}, on higher supergeometry.} The first deliberate forays towards the basis of the subject is \cite{Eder2020} (with application intended to Loop Quantum Gravity \cite{Rovelli2004, Rovelli-Vidotto2014}, rather than supergravity). 

While the pioneers of the (super)group manifold approach to (super)gravity shown clear awareness of Cartan's view \cite{Neeman-Regge1978, Neeman-Regge1978b}, it is less clear for  subsequent contributors. Still, the starting point of the approach is what mathematicans would recognise as Klein supergeometry, i.e.  flat Cartan supergeometry. 
Then, by relaxing the Maurer-Cartan struture equation (switching on curvatures),  one goes from supergroup manifolds to the so-called ``soft supergroup manifolds": the latter corresponds to a Cartan super-bundle. This will be described in section~\ref{Supergroup manifold approach and Cartan supergeometry}.
The supergroup manifold approach can be understood as a version of Cartan supergeometry in the deWitt-Roger approach to supergeometry.
\medskip

In section \ref{Conclusion} we conclude by opening on a further extension of Cartan geometry, to higher (super) geometry, that must be the right mathematical foundation for both supergravity and string theories with higher form (super) potentials  and higher form (super) symmetries.

\section{Classical gauge theories}
\label{Classical gauge theories}

\subsection{Yang-Mills theories: Kinematics and dynamics}
\label{Yang-Mills theories: Kinematics and dynamics}

In this section, we review the geometric underpinning of classical Yang-Mills field theory, as it will provide the background for the discussion the interplay between gauge theories of gravity and Cartan geometry. For another introductory text (aimed at students) see \cite{Francois2021_II} and for more in depth treatments we refer e.g. to \cite{Hamilton2018, Nakahara, Bertlmann}.

\subsubsection{Kinematics = Geometry}
\label{Kinematics=geometry}

The geometry of bundles provide the \emph{kinematics} of a gauge field theory. We briefly summarise this fact via the following dictionary between physics and mathematics.

\medskip
\begin{table}[!h]
\begin{center}
\begin{tabular}{||c | c||}
 \hline
  & \\
 Physics & Mathematics  \\
 [2ex] 
 \hline\hline 
  & \\
 Global/rigid gauge symmetry & Structure group $H$ of a principal bundle $P \xrightarrow{{\small H}} M$  \\[1.5ex]
 Spacetime & $M$ smooth base manifold \\[1.5ex]
 \makecell{Matter fields $\phi$ \\ (particles in QFT)}  & \makecell{ Sections $\Gamma(E)$ of associated bundles $E=P\times_\rho V$ \\ 
 for $H$-representations $(\rho, V)$, $\rho:GH \rarrow GL(V)$ \\
 $\Leftrightarrow\ $ $V$-valued tensorial 0-forms $\vphi\in \Omega^0_{tens}(P, \rho)$ -- (local rep.)}   \\ [4ex] 
 Yang-Mills gauge potential $A$ & (loc. rep.) Ehresmann connection 1-form $\omega$ on $P$  \\[1.5ex]
 Minimal coupling btwn $\phi$ and $A$  & (loc. rep.) Exterior covariant derivative $D\vphi=d\vphi+\rho_*(\omega)\vphi$  \\[1.5ex]
  Yang-Mills field  strength $F$ & (local rep. of) Curvature 2-form  $\Omega \in \Omega_{tens}^2(P, \mathfrak h) $ of $\omega$  \\[1.5ex]
  \makecell{ Sourceless field equation DF=0 \\ 
  {\color{black}(\underline{ex}: Maxwell, Abelian case; dF=0)}} & (local rep. of) Bianchi identity  $D\Omega=0$  \\[3.5ex]
    \makecell{ Gauge transformations/group \\ 
  {\color{black} ( $A \mapsto A'=\gamma^{-1} A \gamma  + \gamma^{-1}d\gamma$} \\
  {\color{black}  $F \mapsto F'=\gamma^{-1} F \gamma$} \\
      {\color{black}  $\phi \mapsto \phi'=\rho(\gamma)^{-1}\phi$\  )} \\[3mm] 
  \phantom{ll} }  & 
\makecell{\emph{Passive GT}: \emph{local gluings} induced by change of local trivialisation \\ via the transition functions of $P$ \\[1.5mm]
   \emph{Active GT}: group of vertical automorphisms of $P$, $\Aut_v(P)$   \\
    $\Leftrightarrow\ $ gauge group of $P$, $\H =\left\{ \gamma: P\rarrow H\ |\ R^*_h\gamma =h^{-1}\gamma h \right\}$  \\
$\Leftrightarrow\ \Gamma (\mathcal{E}) $ with 
$\mathcal E =P \times_{Conj} H$ } \\
\phantom{basta!} & \\
\hline
 \end{tabular}
 \caption{\label{Dico}Bundle geometry and the kinematics of gauge field theories}
 \end{center}
\end{table}

There is much to say about the conceptual and ontological meaning of the presence of gauge symmetries in our most fundamental theories of physics (see e.g. \cite{Berghofer-et-al2023, Francois2018, Rovelli2014}). 
Here, we simply stress that passive gauge transformations are direct analogues of coordinates transformations on $M$ (via its transitions functions), while active gauge transformations are direct analogues of $\Diff(M)$: In GR, coordinate changes are sometimes called ``passive diffeomorphisms", to distinguish from the genuine, ``active", ones. 
The way in which coordinate changes and $\Diff(M)$ encode the fundamental physical insight of GR, the ``relational" character of physics (see \cite{Rovelli2002, Tamborino2012, Francois2023-a}) -- via the \emph{hole and point-coincidence arguments} \cite{Stachel2014, Giovanelli2021} -- is thus mirrored by the way local gluings and $\Aut_v(P)$ encode it in gauge field theory. 
But this obtains only once the dynamics of gauge field theories is understood.  

\subsubsection{Dynamics: Lagrangian and action functionals}
\label{Dynamics: Lagrangian and action functionals}

The  dynamics of a classical field theory is given by field equations, derived from a Lagrangian functional $L$ via the variational principle applied to the corresponding action $S=\int L$.  

The field space of a gauge field theory is $\Phi = \A \times \Gamma(E)$, with $\A$ the space of (loc. rep. of) connections, so that a point is a collection of elementary fields $\upphi=\{A, \phi \}$ or $\upphi=\{\omega, \vphi \}$. The Lagrangian is then
\begin{equation}
\begin{aligned}
L: \Phi \ \ (\text{or } J^1(\Phi) )\ &\rarrow \Omega^m(M) \ \ (\text{or } \Omega^m(P)), \\
\upphi &\mapsto L(\upphi), 
\end{aligned}
\end{equation}
where $J^1(\Phi)$ is the first jet bundle of $\phi$, 
and $m=\text{dim}M$. 
For most physical purpose, it is built as a polynomial in the elementary field variables $\phi$ 
-- and their Hodge duals, the Hodge operator $\, *:\Omega^p(M) \rarrow \Omega^{p-m}(M)$ being a natural tool to form top forms on $M$.

The action is thus
\begin{equation}
\begin{aligned}
S: \Phi \ \ (\text{or } J^1(\Phi) )\ &\rarrow \RR, \\
\upphi &\mapsto S(\upphi)=\int_D L(\upphi), 
\end{aligned}
\end{equation}
with $D \subset M$ a domain of spacetime. Physically realised field configurations are those that make the action stationary $\delta S =0$. 
The variation of the Lagrangian gives
\begin{equation}
\begin{aligned}
\delta L = \bs E + d\bs\theta \defeq E(\delta \upphi; \upphi) + d\theta(\delta \upphi; \upphi),
\end{aligned}
\end{equation}
where $\bs E= E(\delta \upphi; \upphi)$ is the field equation term. The boundary term $\bs \theta=\theta(\delta \upphi; \upphi)$ -- sometimes called the presymplectic potential of the theory -- has vanishing contribution if $\d D=\emptyset$, or given adequate boundary conditions (b.c.) on $\upphi$:\footnote{Neumann b.c.  are $\upphi_{|\d D}=0$, Dirichlet b.c. are $\delta \upphi_{|\d D}=0$.} Either conditions are  required 
if we are to get the field equations of $\upphi$ from the stationary action principle: $\delta S=0\ \ \Rightarrow \ \ \bs E=0$, for all $\delta \upphi$. 
\medskip

The choice of Lagrangian is usually  constrained by \emph{symmetry principles}: $L$ is required to be invariant (up to boundary terms) under some continuous transformation groups. This ensures that the field equations derived from it are \emph{covariant} under these groups.
The two principles at the heart of relativistic gauge field theory are the General Covariant Principle (GPC), or General Relativity Principle, and the Gauge Principle (GP).

The GCP requires that physical laws (equations) are invariant or covariant under arbitrary changes of coordinates. It can be satisfied easily by working with tensors and differential forms. 

The GP requires that $L(\phi)$ has trivial gluings on $M$, i.e. it is invariant under passive gauge transformations, or equivalently is \emph{basic} on $P$ (invariant and horizontal) -- $L(\phi) \in \Omega^m_\text{bas}(P)$ -- meaning it is invariant under active gauge transformations $\H \simeq \Aut_v(P)$. To satisfy the GP, one may build $L(\phi)$ using 1) tensorial forms $\Omega^\bullet_\text{tens}(P, \rho)$ and 2) $H$-invariant non-degenerate bilinear forms (or polynomials) 
\begin{itemize}
    \item on $\LieH$: the Killing form, or the trace, $k(\ ,\ )=\Tr : \LieH \times \LieH \rarrow \RR$;
    \item on $V$: an inner product, $\langle\ ,\ \rangle: V \times V \rarrow \RR$.
\end{itemize}
\medskip
We remark that the non-degeneracy condition is essential if $E(\delta \upphi; \upphi)=0$ $\forall \delta \upphi$ is to give the field equation as desired. 
Let us consider a few typical examples: 
\medskip

\noindent $\bullet$ The Yang-Mills Lagrangian, for $H=S\!U(n)$, is
\begin{align}
    L_\text{YM}(A) = \tfrac{1}{2}k(F, *F)
                   = \tfrac{1}{2}\Tr(F *\!F).
\end{align}
It specifies the dynamics of the Yang-Mills potential $A$. The kinematic term comes from the derivative of $A$ in $F$, while the bracket term $[A,A]$ in $F$ shows that $A$ is self-interacting:\footnote{It carries a YM-charge, or ``color" charge -- by analogy with Quantum ChromoDynamics (QCD), the gauge theory of the strong nuclear force -- to which it is sensible.}
the non-linearity of YM theory is due to the non-Abelian nature of the underlying group.
The special Abelian case $H=U(1)$ gives the Maxwell Lagrangian $L_\text{Max}(A)=\tfrac{1}{2} F *F$, a linear theory where there is no self-interaction.\footnote{The electromagnetic field (thus photon in QFT) does not carry electric charge.}
Observe that a mass term for $A$, which must be of the form $\,m\Tr(A *A)$, is not gauge-invariant: gauge symmetries compels YM fields to be massless.
\medskip

\noindent $\bullet$ The Klein-Gordon Lagrangian is
\begin{align}
    L_\text{KG}(\phi) = \langle D\phi, *D\phi\rangle
                        + \mu^2 \langle \phi, *\phi\rangle.
\end{align}
It describes the dynamics of a $S\!U(n)$ scalar field $\phi$, with mass $\mu$, minimally coupled with an external (i.e. without dynamics) Yang-Mills potential $A$.
\medskip

\noindent $\bullet$ The Dirac Lagrangian is
\begin{align}
L_\text{Dirac}(\psi)= \langle\psi,\slashed{D}\psi\rangle
                    - \mu \langle \psi, *\psi\rangle.
\end{align}
It describes the dynamics of a $S\!U(n)$-charged Dirac spinor $\psi\in \Gamma(\S \otimes E)$ -- $\S$ a spin bundle over $M$ -- with mass $\mu$,  minimally coupled with an external YM potential $A$. Such spinors describe fundamental matter (fermionic) fields. The Dirac operator $\slashed D= \upgamma \wedge *D$ involves $\upgamma =\upgamma_\mu dx^{\,\mu} = \upgamma_a e^a = \upgamma_a {e^a}_\mu dx^{\,\mu}$, with $e^a$ the (local) soldering form --  known in physics as the tetrad field or \emph{vierbein} --  and $\upgamma_a$ Dirac gamma matrices. 
\medskip

\noindent $\bullet$ The prototypical Lagrangian for models displaying a phenomenon of ``spontaneous gauge symmetry breaking" is
\begin{align}
L_\text{SSB}(A, \phi)= 
\tfrac{1}{2}\Tr(F *F)
\ +\ \langle D\phi, *D\phi\rangle +  V(\phi),
\end{align}
where the potential term $V(\phi)$ for the scalar $\phi$ 
-- often called a Higgs field in this context -- 
is at most a quadratic polynomial 
$V(\phi)=\alpha \langle \phi, *\phi\rangle + \beta \langle \phi, *\phi\rangle^2$. 
For adequate sign of the parameters $\alpha, \beta$ the vacuum expectation value (VEV) $\phi_0$ of the scalar, i.e. solution of ${\delta V(\phi)}/{\delta \phi}=0$, is  non-zero. By expanding the scalar as $\phi=\phi_0 + \sf{H}$  in the above Lagrangian,  mass terms for $\sf{H}$ 
\emph{and} the YM field $A$ depending on $\phi_0$ emerge. 
Such a model, allowing to endow YM fields with mass despite gauge symmetries, features prominently in the Lagrangian  of the Standard Model of particle physics, describing the ElectroWeak sector (with group $U(1)\times S\!U(2)$).
It includes a Dirac Lagrangian $L_\text{Dirac}(\psi)$ where  $\phi_0$-dependent mass terms for matter spinors emerge from the so-called Yukawa couplings between $\psi$ and $\phi$: the scalar/Higgs field endows any field it interacts with, gauge and matter fields, with mass (depending on the VEV $\phi_0$ and the strength of the specific gauge and Yukawa coupling constants).

\medskip
We remark that Lagrangians that do not involve the Hodge operator (or the soldering form), thus do not rely on a metric structure on $M$, are called ``topological theories". A well-known case is the 3-dimensional $S\!U(n)$ Chern-Simons Lagrangian 
\begin{align}
L_\text{CS}(A)= \Tr\left(AdA+ \tfrac{2}{3}A^3 \right).
\end{align}
We observe that it is not gauge-invariant; yet, as we will see, the field equation remains gauge-covariant: from $\delta L_\text{CS}$ one indeed find $\bs E =E(\delta A; A)= \Tr(\delta A\, F)=0$ for all $\delta A$, i.e. $F=0$.  

\subsubsection{Noether charges and gauge symmetries}
\label{Noether charges and gauge symmetries}

Noether theorems  express that variational symmetries of the Lagrangian and action functionals give physical quantities, known as charges, conserved on-shell, i.e. when the field equations hold. Since $L$ is gauge-invariant, we expect charges to abound in gauge-field theory. 

Denote $\chi\in$ Lie$\H$ any element of the Lie algebra of the gauge group and $\delta_\chi$ the corresponding infinitesimal gauge transformation operator.\footnote{If we are on the bundle $P$, it is $L_{\chi^v}$, the Lie derivative along the vertical vector field $\chi^v \in \Gamma(VP)$ generated by $\chi$. }
As a special case of the variational principle one gets 
\begin{equation}
\begin{aligned}
\delta_\chi L =  
E(\delta_\chi \upphi; \upphi) + d\theta(\delta_\chi \upphi; \upphi).
\end{aligned}
\end{equation}
According to the GP, $\delta_\chi L=0$, so the quantity $J(\chi; \upphi)\defeq \theta(\delta_\chi \upphi; \upphi)$ is $d$-closed on-shell, which we denote 
$dJ(\chi; \upphi)\ \hat{=}\ 0$. 
It is the Noether $(m-1)$-form current. 
It is thus on-shell a boundary term: $J(\chi; \upphi)\ \hat{=}\ d q(\chi; \upphi)$. 
The corresponding Noether charge is its integral over a codimension 1 submanifold $\Sigma\subset D$,
\begin{equation}
\begin{aligned}
Q_\Sigma(\chi; \upphi)\defeq \int_\Sigma J(\chi; \upphi)=\int_\Sigma \theta(\delta_\chi \upphi; \upphi)
\ \hat{=}\ \int_{\d\Sigma} q(\chi; \upphi).
\end{aligned}
\end{equation}
By the last equality, the Noether charge is a boundary term. For this reason it is often called a ``quasi-local " charge, as it is a quantity associated to a region, not a single point, of spacetime.  

Let us consider the notable examples of YM  theory. Variation of its Lagrangian gives
\begin{equation}
\begin{aligned}
\delta L_\text{YM} 
&=  E(\delta A; A) + d\theta(\delta A; A) \\
&= \Tr(\delta A\ D\!*\!F ) + d\Tr(\delta A *\!F). 
\end{aligned}
\end{equation}
From this one reads the (vacuum) YM equation $\bs E=0 \ \Rightarrow\  D*\!F=0$. The Abelian case is the  Maxwell equation $d*\!F =0$.
Given the infinitesimal gauge transformation of the YM potential, $\delta_\chi A= D\chi= d\chi+ [A, \chi]$, the Noether charge is
\begin{align}
Q^\text{YM}_\Sigma(\chi; A)
=\int_\Sigma \theta(\delta_\chi A; A)
=\int_\Sigma \Tr(D\chi *\!F)
\ \, \hat{=} \int_{\d\Sigma} \Tr(\chi *\!F).
\end{align}
For constant $\chi$,  i.e. for elements of the Lie algebra of the structure group of the underlying bundle $P$, this gives a non-Abelian Gauss law \cite{Abbott-Deser1982, Francois-et-al2021}. Indeed, in the Abelian case, $Q^\text{YM}_\Sigma(\chi; A)= \chi \int_{\d\Sigma}(*F)$, the integral is the Gauss law: the electric flux through a bounded region which gives the enclosed electric charge. 
\medskip





\subsection{Classical gauge theory of gravity and Cartan geometry}
\label{Classical gauge theory of gravity and Cartan geometry}

In this section we review  elementary notions of standard Cartan geometry and show how they apply naturally to the gauge  theoretic description of gravitational physics. For in depth treatments, one should consult the reference texts \cite{Cap-Slovak09, Sharpe, Kobayashi1972}. See also \cite{Kobayashi1957} which is of historical interest.


\subsubsection{Cartan geometry in a nutshell}
\label{Cartan geometry in a nutshell}

A Cartan geometry $(P, \varpi)$ is a 
$H$-principal bundle $P$ endowed with a Cartan connection $\varpi \in \Omega^1(P, \LieG)$, with $\LieG \supset \LieH$, s.t. 
\smallskip

\begin{enumerate}[label=\roman*, leftmargin=*, itemsep=3pt]
\item $R^*_h \varpi_{|ph}=\Ad_{h\-} \varpi_{|p}$,
\item $\varpi_{|p}(X^v_p)=X \in$ $\LieH$, where $X^v_p \in V_p\P$,
\item $\varpi_{|p} : T_p P \rarrow$ $\LieG\ $ is a linear isomorphism $\forall p\in \P$. 
\end{enumerate}
\smallskip

\noindent Properties i-ii are the same as for an Ehresmann connection. The last property iii is key and has several consequences special to Cartan geometry. The essential idea, from a physical standpoint, is that it implies $P$ encodes the geometry of $M$. Since the Einsteinian insight is that gravity is the geometry of spacetime,  Cartan geometry $(P, \varpi)$ is perfectly adapted to describe the kinematics of gauge theories of gravitation, the Cartan connection being a generalised gravitational gauge potential. Let us see how. 
\medskip

\noindent {\bf Gauge symmetries}. 
As gauge approaches to gravity were often modeled on the habits of YM theory, whenever one saw a $\LieG$-valued  1-form, understood as a gauge potential, it was assumed that the associated theory should have a gauge symmetry $\G$ -- or gauge algebra $\LieG$ -- corresponding to the Lie algebra $\LieG$.  
This led often to technical complications and conceptual issues to be dealt with. For example, in the case $\LieG = \LieH \ltimes \RR^4$ typical of affine gravity ($\LieH=\gl(4)$) or Poincaré gravity ($\LieH=\so(1,3)$), the thorny issue of ``internal gauge translations" and their redundancy with diffeomorphisms $\Diff(M)$ needed resolution. Mostly the strategy has been to try to identify ``gauge translations" with $\diff(M)$ (via the tetrad field/soldering form) \cite{Blagojevi-et-al2013}.

No such problem arises once one knows Cartan geometry. The underlying bundle being a $H$-bundle $P$, the gauge group is $\H$, i.e. the gauge transformations result from the action of its vertical automorphisms group $\Aut_v(P)$. All the relevant symmetries of gravity, including $\Diff(M)$, are then naturally organised in the short exact sequence (SES) of groups associated to the Cartan bundle,
 \begin{align}
 \label{SESgroup}
\H \simeq \Aut_v(P)  \rarrow   \Aut(P) \rarrow \Diff(M).
 \end{align}
The corresponding SES of Lie algebras is
 \begin{align}
 \label{SESalg}
Lie\H \simeq \aut_v(P)  \rarrow   \aut(P) \rarrow \diff(M),
 \end{align}
where $Lie\H$ is the `gauge algebra' and $\aut(P)$ is isomorphic to $\Gamma_\text{inv}(TP)$, the right-invariant vector fields of $P$. The sequence \eqref{SESalg} describes the Atiyah Lie algebroid of $P$, and encodes the infinitesimal symmetries or gravity \cite{Attard-et-al2019}. 
\bigskip

\noindent {\bf Soldering}.
Given its defining properties, the Cartan connection $\varpi$ induces a soldering on $M$, i.e. the bundle isomorphism $TM \simeq P\times \LieG/\LieH$.\footnote{Unless $\LieH$ is  an ideal of $\LieG$, $\LieG/\LieH$ is a vector space, not a subalgebra. In any case, $H$ acts on it via the $\Ad$ representation, which may reduce to simple left multiplication in some cases. As for $\LieG$, it acts via the $\ad$ representation.}  
So $\Gamma(TM) \simeq \Omega^0_\text{eq}(P, \LieG/\LieH)$.
Conceptually, it shows how $P$ encodes the geometry of $M$: its tangent bundle being naturally an associated bundle of $P$. 

Further consequences unfold, that are relevant for a field theoretic standpoint. 
Consider the projection $\tau: \LieG \rarrow \LieG/\LieH$. One defines the soldering form $\theta\defeq \tau\circ \varpi\, \in \Omega^1_\text{tens}(P, \LieG/\LieH)$. It is tensorial indeed; equivariant by property i and horizontal by ii.
Locally, on an open set $U\subset M$, given a trivialising section $\s:U\rarrow P_{|U}$, the local representative of the Cartan connection is $\b A\defeq \s^* \varpi \, \in\Omega^1(U, \LieG)$, while the local soldering form is
$e \defeq \s^* \theta \, \in\Omega^1(U, \LieG/\LieH)$. 
Given a coordinate system $\{x^\mu\}$ on $U$ and $\{a\}$ abstract index for $\LieG/\LieH$, we may write $e=e^a={e^a}_\mu\, dx^{\,\mu}$. The components ${e^a}_\mu$ of the local soldering are called \emph{tetrad field}, or \emph{vielbein}, in physics. 

If there is a non-degenerate bilinear form  $\eta: \LieG/\LieH \times \LieG/\LieH \rarrow \RR$, then the Cartan connection induces a metric via
\begin{equation}
 \begin{aligned}
 \label{metric}
g\defeq \eta \circ e : \Gamma(TM) \times  \Gamma(TM) &\rarrow C^\infty(M), \\
(X, Y) & \mapsto g(X, Y)=\eta\left(e(X), e(Y) \right).
 \end{aligned}
 \end{equation}
In coordinates this reads
$g_{\mu\nu}=\eta_{ab} {e^a}_\mu{e^b}_\nu$.
Under another local section $\s'=R_g \s$, with $g:U\rarrow H$ a transition function of $P$, $e'=\s'^*\theta=R^*_g e=\Ad_{g\-} e$.
If $\eta$ is $H$-invariant, $\eta \circ \Ad(H) =\eta$, then $g'=\eta\circ e' = \eta\circ e =g$: the induced metric is well-defined on $M$. It thus allows to define a Hodge operator $*:\Omega^p(M) \rarrow \Omega^{p-m}(M)$, an essential ingredient to built Lagrangians of 
 (non-topological) gauge field theories (even non-gravitational), as we have seen.

One may observe that, in the case $(\LieG/\LieH, \eta)=(\RR^4, \eta)=\sf M$ the Minkowski space,  the existence of the local soldering $e\in\Omega^1(U, \RR^4)$, and the expression  \eqref{metric},   are mathematical implementation of one formulation of the \emph{equivalence principle}: Spacetime is locally Minkowskian, and it is always possible to find a coordinate system in which the metric appears Minkowskian, i.e. spacetime appears locally flat. 

In standard metric-formulation of GR, the metric is often understood as the gravitational potential (as the Newtonian potential in the weak field limit). We also observe that the $\LieH$-part $\omega$ of $\varpi$ is Ehresmann by properties i-ii, i.e. an exact analogue of Yang-Mills potential in YM gauge field theory. So, the Cartan connection $\varpi$ (or rather its local representative $\b A$) is  a generalised \emph{gravitational gauge potential}. 
\bigskip

\noindent {\bf Cartan curvature}.
The curvature 2-form of a Cartan connection is \emph{defined} via the Cartan structure formula: 
\mbox{$\b\Omega=d\varpi+\sfrac{1}{2}[\varpi, \varpi]\, \in \Omega^2(P, \LieG)$}. It thus satisfies the Bianchi identity $d\b\Omega + [\varpi, \b\Omega]\equiv 0$, a strictly algebraic identity. 
The torsion of a Cartan connection is 
$\Theta = \tau \circ \b\Omega\, \in \Omega^2(P, \LieG/\LieH)$. This is typical of what Cartan called the ``espaces généralisés". This is a notion of course absent for Ehresmann connections, thus in YM theory. 
Locally, on $U\subset M$, we have the local representatives $\b F = d\b A + \sfrac{1}{2}[\b A, \b A]\, \in \Omega^2(U, \LieG)$ and 
$T \in \Omega^2(U, \LieG/\LieH)$. 

In direct anology to gauge field theory, where the YM field strength is the curvature of an Ehresmann connection (Table \ref{Dico}), the Cartan curvature (its local representative) is the \emph{gravitational field strength}.
\bigskip

\noindent {\bf Associated (tractor) bundles and covariant (tractor) derivative}.
As usual, given representations $(\rho, V)$ of $H$, one have associated bundles 
$E=P\times_\rho V$ and $\Gamma(E)\simeq \Omega^0_\text{eq}(P, V)$. 
If $V$ is also a $\LieG$-module, with $\rho_*:\LieG \rarrow \gl(V)$, 
on account of i-ii, $\varpi$ induces a covariant derivative 
$\b D=d\ + \b\rho_*(\varpi) : \Omega^\bullet_\text{tens}(P, \rho) \rarrow 
\Omega^{\bullet+1}_\text{tens}(P, \rho)$. The Bianchi identity can thus be written $\b D\b \Omega\equiv0$.
It is easily seen that $\b D \circ \b D = \rho(\b\Omega)$. In view of Table \ref{Dico}, one can see that $\b D$ will represent the \emph{minimal coupling to gravity}. 

For $(\b\rho, V)$ a representation of $G$ (the group with Lie algebra $\LieG$), which is a representation of $H$ by restriction, one defines tractor bundles: $\T \defeq P\times_{\b\rho} V$.
The covariant derivative acting on $\Gamma(\T)\simeq \Omega_\text{tens}(P, \b\rho)$ is then called a \emph{tractor derivative}, or \emph{tractor connection}. An important example is, for $(\Ad, \LieG)$, the adjoint tractor bundle $\A M \defeq P \times_\Ad \LieG$. The Cartan curvature can be seen as a 2-form on $M$ valued in $\A M$. 

When $H$ or $G$ are $S\!O$ groups, and $\tilde\rho: G \rarrow \tilde G$ are their spin lifts (representations), one may define spin bundles $\S \defeq P\times_{\tilde\rho} \t V$ and spinor fields $\Gamma(\S) \simeq \Omega_\text{tens}(P, \t\rho)$.  
For example, when $H=S\!O(1,3)$, so $\t H=S\!L(2, \CC)$, $\Gamma(\S)$ are Dirac spinors describing matter fields (fermions), as we have recalled earlier, and $\b D$ their coupling to gravity. For $G=S\!O(2,4)$, so $\t G=S\!U(2,2)$, $\Gamma(\S)$ are twistor fields (à la Penrose) and $\b D$ is the twistor connection. 
\bigskip

\noindent {\bf Normal Cartan connection}.
The notion of normal Cartan connection, or normal Cartan geometry, generalises the idea of Levi-Civita connection in (pseudo-)Riemannian geometry. The idea, at least for physically interesting geometries (e.g. reductive or parabolic, see below), is to find a Cartan connection expressed uniquely in terms of the soldering part. 
One thus imposes (algebraic and/or differential) constraints on $\varpi$ so as to express its $\LieH$-part $\omega$ in terms of its $\LieG/\LieH$-part $\theta$.  
Those~constraints should be gauge-invariant, so they must be imposed on $\b\Omega$. 
The unique normal Cartan connection is thus $\varpi_{\text{{\tiny N}}}=\varpi_{\text{{\tiny N}}}(\theta)$.

For example, in Cartan-Riemann or Cartan-de Sitter (that we will see shortly) it is enough to impose the vanishing of the $\LieG/\LieH$-part of $\Omega$, i.e. vanishing torsion $\Theta =0$. In these cases, the $\LieH$-part $\omega=\omega(\theta)$  of $\varpi_{\text{{\tiny N}}}$ is  the usual Levi-Civita connection.  
In the class of parabolic geometries, defined in the next section, there is a nice characterisation of normal Cartan connection in terms of the Kostant-Spencer cohomology of the parabolic algebra $\LieG$ -- see \cite{Cap-Slovak09, Ochiai, Cap-et-al1997}. As an example, we will give the normality conditions in the conformal Cartan geometry in the next section.
\medskip

Physically, in vacuum General Relativity or for non-spinorial matter sources (scalar fields, like fluids, or dust as in cosmology), the gravitational d.o.f. are those of the metric, or the soldering form. The normality conditions thus eliminate the extra d.o.f. in the $\LieH$-part of the Cartan connection, so that $\varpi_{\text{{\tiny N}}}$ contains precisely the gravitational d.o.f.: the normal Cartan connection is exactly the gravitational potential. 
In a more  fundamental description,    matter  sources are  described by spinor fields $\psi \in \Gamma(\S)\simeq \Omega^0_\text{tens}(P, \t\rho)$, and their coupling to gravity is given by $\b D\psi=d\psi +\t\rho_*(\varpi)$ (which contains the Dirac operator). The spinor field sources both the Einstein tensor and the torsion tensor: the Cartan connection, as the gravitational potential, thus cannot be normal.

\subsubsection{Special cases and application to physics}
\label{Special cases and application to physics}

\medskip

\noindent {\bf Klein geometry}.
Consider a Lie group $G$ and a closed subgroup $H$ s.t. $G/H$ is an homogeneous space.  
A Klein geometry\footnote{After the mathematician Felix Klein, who suggested in his 1872 Erlangen program to classify and study non-Euclidean homogeneous spaces via their transformation groups. }  is $(G, \varpi_{\text{{\tiny G}}})$ where $G$, called the principal group, is seen as a principal $H$-bundle $G \xrarrow{H} G/H$, and $\varpi_{\text{{\tiny G}}} \in \Omega^1_\text{tens}(G, \LieG)$ is its Maurer-Cartan form. As a special case of its definition, the latter satisfies
\smallskip

\begin{enumerate}[label=\roman*, leftmargin=*, itemsep=3pt]
\item $R^*_h \varpi_{\text{{\tiny G}}}=\Ad_{h\-} \varpi_{\text{{\tiny G}}}$,
\item $\varpi(X^v)=X \in$ $\LieH$, where $X^v_p \in V_p\P$,
\item ${\varpi_{\text{{\tiny G}}}}_{|p} : T_p\P \rarrow$ $\LieG\ $ is a linear isomorphism $\forall p\in \P$. 
\end{enumerate}
\smallskip

\noindent 
One says that $(G, H)$ is the Klein pair on which a Klein geometry is based. 
For example, standard Euclidean or Minkowskian geometries are special cases of Klein geometries based respectively on the pair $\left( S\!O(n) \ltimes \RR^n, S\!O(n)\right)$ and $\left( S\!O(1, n-1) \ltimes \RR^n, S\!O(1, n-1)\right)$.

In addition, ${\varpi_{\text{{\tiny G}}}}$ satisfies the Maurer-Cartan equation $d\varpi_{\text{{\tiny G}}} +\tfrac{1}{2}[\varpi_{\text{{\tiny G}}}, \varpi_{\text{{\tiny G}}}] =0$. 
The Maurer-Cartan form is thus a flat Cartan connection on the bundle $G$, and  Klein geometries $(G, \varpi_{\text{{\tiny G}}})$ are flat Cartan geometries.
One often says that $(G, H)$ is the Klein pair modeling a Cartan geometry $(P, \varpi)$. And a general flat Cartan geometry is (locally) isomorphic to a Klein geometry: $(P, \varpi)_{\text{{\tiny flat}}} \simeq (G, \varpi_{\text{{\tiny G}}})$, which imply that the base manifold is homogeneous $M\simeq G/H$. 
Flatness in the sense of Cartan therefore generalises flatness in the sense of (pseudo-)Riemannian~geometry. 
We~reproduce here the nice  diagram of Sharpe \cite{Sharpe}, 
\begin{align}\label{Cartan geometry}
\makebox[\displaywidth]{
\hspace{-18mm}
\begin{tikzcd}[column sep=normal, row sep=scriptsize, ampersand replacement=\&]
\&           \text{Euclide (Minkowski)}              \arrow[d]  \arrow [r]                  \&               \text{ Riemann (Lorentz)   }           \arrow[d]        \\
  \&           \text{ Klein }                      \arrow[r, ]                                              \&    \text{ Cartan   }  
\end{tikzcd}}  \raisetag{5.7ex}
\end{align}
which illustrates how, in a beautiful interplay of group theory and differential geometry, Cartan geometry is the common generalisation of Klein and Riemann geometries.
\bigskip

\smallskip

\noindent {\bf Reductive and parabolic Cartan geometries}.
The subclass of  reductive  Cartan geometries is especially noteworthy: they are those for which there is an $\Ad(H)$-invariant decomposition $\LieG=\LieH \oplus V$. This means we have a clean split of the Cartan connection as $\varpi= \omega+ \theta$, where $\omega$ is a Ehresmann connection on $\P$. The curvature splits accordingly $\b \Omega=\Omega+\Theta$. Pseudo-Riemannian  geometry,\footnote{As reformulated by Cartan via his ``moving frame'', and independently by Einstein via his ``vierbein/vielbein'', i.e. the soldering $e$.} which are based on $\mathfrak{iso}(r,s)=\so(r,s)\oplus\RR^m$, belongs to this subclass. 

Parabolic Cartan geometries are another remarkable subclass where one has a $|k|$-grading of $\LieG$, 
  i.e. $\LieG=\bigoplus_{-k\leq i \leq k} \LieG_i$ s.t. $[\LieG_i, \LieG_j] \subset \LieG_{i+j}$, and $H$ is s.t. $\LieH=\bigoplus_{0 \leq i \leq k} \LieG_i$.  Both $\varpi$ and $\b\Omega$ split along the $|k|$-grading, and here also the Lie$H$-part of $\omega$ is a Ehresmann connection. An important example is conformal Cartan geometry, based on the $|1|$-graded algebra $\so(r+1, s+1)=\RR^m \oplus \co(r, s) \oplus \RR^{m*}$, with $\co(r, s)=\so(r, s) \oplus \RR$, and where $H$ is s.t. $\LieH=\co(r, s) \oplus \RR^{m*}$. The spin version in case $m=4$, based on the $|1|$-grading of $\su(2, 2)$, is very closely related to twistor geometry \cite{Penrose-Rindler-vol1, Penrose-Rindler-vol2}. 

Let us consider three notable examples, two reductive geometries and one parabolic, together with related physical applications to gravity theories. 
\bigskip

\noindent $\bullet$ {\bf Cartan-Riemann geometry}:
This geometry $(P, \varpi)$ is based on the  group $G=H\ltimes \RR^m = S\!O(r,s)\ltimes \RR^m
=\begin{psmallmatrix} S & 0 \\ 0 & 1 \end{psmallmatrix}
 \begin{psmallmatrix} 0 & t \\ 0 & 1 \end{psmallmatrix}$, where we use a matrix notation to make computations easy. 
 The homogeneous (flat) model is $G/H=\RR^m$, equipped with the metric $\eta$ preserved by $H=S\!O(r,s)$.  
 The Cartan connection and its local representative are
 \begin{equation}
 \begin{aligned}
 \label{Cartan-co-CR}
\varpi= \begin{pmatrix} \omega & \theta \\ 0 & 0 \end{pmatrix}
\quad \text{ and } \quad 
\b A= \begin{pmatrix} A & e \\ 0 & 0 \end{pmatrix},
 \end{aligned}
 \end{equation}
 while the Cartan curvature and its local representative are
 \begin{equation}
 \begin{aligned}
  \label{Cartan-curv-CR}
\b\Omega= \begin{pmatrix} \Omega & \Theta \\ 0 & 0 \end{pmatrix}
\quad \text{ and } \quad 
\b F= \begin{pmatrix} R & T \\ 0 & 0 \end{pmatrix}
=\begin{pmatrix} dA+ \tfrac{1}{2}[A, A] & de+Ae \\ 0 & 0 \end{pmatrix},
 \end{aligned}
 \end{equation}
 where $R$ is the Riemann curvature 2-form. 
The normality condition is just $\Theta=0$, so that the normal Cartan connection is $\varpi_{\text{{\tiny N}}}=\omega(\theta) + \theta$, with $\omega(\theta)$ the Levi-Civita connection -- in this form, sometimes known as the ``spin connection". 
  
 The gauge group of the bundle is $\Aut_v(P)\simeq \SO(r,s)$, i.e. the \emph{local} pseudo-rotation group, with element 
 $\gamma =\begin{psmallmatrix} \sS & 0 \\ 0 & 1 \end{psmallmatrix}$.
 The gauge transformations of the local Cartan connection and its curvature are
 \begin{equation}
 \begin{aligned}
{\b A}^\gamma&=\gamma\- \b A \gamma + \gamma\-d\gamma 
   &\quad \text{ and } \qquad
{\b F}^\gamma&=\gamma\- \b F \gamma
\\
&=\begin{pmatrix} \sS\- A \sS + \sS\-d\sS\  & \sS\- e \\ 0 & 0 \end{pmatrix}
& \quad
&=\begin{pmatrix} \sS\- R \sS & \sS\-T \\ 0 & 0 \end{pmatrix}.
 \end{aligned}
 \end{equation}
The linear transformation of the connection, for $\chi =\begin{psmallmatrix} \chi & 0 \\ 0 & 0 \end{psmallmatrix} \in$ Lie$\SO(r,s)$, is $\delta_\chi \b A=\b D\chi=d\chi +[\b A, \chi]$, splitting as $\delta_\chi A =D\chi =d\chi + [A, \chi]$ and $\delta_\chi e = -\chi e$. 

Given the $S\!O(r, s)$-invariant bilinear form $\eta: \RR^m \times \RR^m \rarrow \RR$, we have an induced metric $g=\eta \circ e$ on $M$. It allows to defined a Hodge operator on $\Omega^\bullet(M)$. 
\medskip

\noindent \underline{{\bf Physics}}: The above geometry is usually taken to provide the kinematics of a formulation of General Relativity known variously as the tetrad formulation, the Palatini formulation, the Sciama-Kibble formulation, or yet, the Einstein-Cartan formulation. It is the following. 

If one is to built a gauge theory of gravitation based on the above geometry, for $m=4$ and $(r, s)=(1,3)$ -- i.e. the gauge symmetry of the theory is the local Lorentz group -- its Lagrangian 
\begin{equation}
\begin{aligned}
L:\Phi=\b \A &\rarrow \Omega^4(M),\\ 
\b A &\mapsto L(\b A)=L(A, e) 
\end{aligned}
\end{equation}
must be $\SO(1,3)$-invariant to satisfy the gauge principle. 
One may use the two immediately available tensorial form, $e$ and $\b F$. Given the $S\!O$-invariant polynomial $M \bullet N\defeq M^{ab}N^{cd}\epsilon_{abcd}$, one may feed it the antisymmetric 2-forms $e \w e^T$ and $R \eta\-$, with $\eta$ the Minkowski metric on $\R^4$. 
The Lagrangian of vacuum GR (without matter) with cosmological constant $\Lambda$ is
\begin{equation}
\begin{aligned}
L_{\text{{\tiny EC}}}(\b A) &= R \eta\- \bullet e \w e^T - \tfrac{\Lambda}{6} e \w e^T \bullet e \w e^T \\
&= \left(R^{ab} e^ce^d - \tfrac{\Lambda}{6} e^a e^b e^c e^d\right) \epsilon_{abcd}.
\end{aligned}
\end{equation}
The cosmological constant $\Lambda$ is  known to be non-zero and positive since a SnIa survey in 1998. 
The variational principle gives
\begin{equation}
\begin{aligned}
\delta L_{\text{{\tiny EC}}}  &= \bs E_{\text{{\tiny EC}}}+d \bs\theta_{\text{{\tiny EC}}} = E(\delta \b A; \b A)+ d\theta(\delta \b A; \b A) \\
&= 2\left(\delta e \w e^T \bullet (R - \tfrac{\Lambda}{3} e \w e^T) + \delta A \bullet T \bullet e^T \right)
\ + \ d \left(\delta A \bullet e \w e^T \right).
\end{aligned}
\end{equation}
The piece of $\bs E_{\text{{\tiny EC}}}=0$ linear in $\delta e$ is Einstein's field equation. The piece linear in $\delta A$ is the torsion-free condition in vacuum, implying $A=A(e)$: i.e. the normal Cartan connection is solution of the vacuum field equation. 
From the potential one get the $\SO(1,3)$-Noether charge 
\begin{equation}
\begin{aligned}
Q_\Sigma^{\text{{\tiny EC}}}(\chi; \b A)=
\int_\Sigma \delta_\chi A \bullet e\w e^T
= \int_\Sigma D\chi  \bullet e\w e^T 
\ \hat{=}\  \int_{\d \Sigma} \chi  \bullet e\w e^T.
\end{aligned}
\end{equation}
The last equality, noted on-shell, actually uses only $T=0$, not Einstein's equation. 
In the theory coupled to matter, $L_{\text{{\tiny EC}}}(\b A)+L_{\text{{\tiny matter}}}(\phi)$, this charge essentially gives the mass of  the distribution of  source matter enclosed by $\d\Sigma$.\footnote{When $\chi$ is the gradient of a timelike Killing symmetry ($\chi^{ab}=\d^a\xi^b$), the charge is known as the Komar mass, or Komar integral. See e.g. \cite{Francois2021, Kastor2008, Choquet-Bruhat2009}.} 

One may observe that the groundstate of the theory is the de Sitter (or anti-de Sitter) space $M=(A)dS$, and not the homogeneous model, the Minkowski space ${\sf M}=(R^4, \eta)$. One also see that the charge is a priori non-zero on the groundstate, which is physically unappealing (given the above interpretation). 
So, the physical theory is misaligned with the underlying geometry providing its kinematics. These defects are cured by selecting the right Cartan geometry, which we discuss next. 
\bigskip

\noindent $\bullet$ {\bf Cartan-de Sitter geometry}:
This geometry $(P, \varpi)$ is based on the  group $G=S\!O(1,4)$ (or $G=S\!O(2,3)$) and  $H= S\!O(1,3)$, so that 
the homogeneous models is the (anti-)de Sitter space, $G/H\simeq (A)dS$. The gauge group is thus still the local Lorentz group, $\Aut_v(P)\simeq \SO(1,3)\ \ni\ \gamma =\begin{psmallmatrix} \sS & 0 \\ 0 & 1 \end{psmallmatrix}$. 
The Lie algebra splits as $\LieG=\LieH \oplus \RR^4$, with elements written in matrix form as
$\begin{psmallmatrix} s & \tau \\ -\epsilon\tau & 0 \end{psmallmatrix}$, where $\tau^t=\tau^T \eta$, $\eta$ being the Minkowski metric, and $\epsilon=\pm$ for the $dS$ and $AdS$ cases, respectively. 
We may thus write the Cartan connection and its local representative as
 \begin{equation}
 \begin{aligned}
\varpi= \begin{pmatrix} \omega & \tfrac{1}{\ell}\theta \\ \tfrac{-\epsilon}{\ell}\theta^t & 0 \end{pmatrix}
\quad \text{ and } \quad 
\b A= \begin{pmatrix} A & \tfrac{1}{\ell}e \\ \tfrac{-\epsilon}{\ell}e^t & 0 \end{pmatrix},
 \end{aligned}
 \end{equation}
 with $\ell$ a constant with dimension of length s.t. $\tfrac{1}{\ell^2} =\tfrac{\Lambda}{3}$. The Cartan curvature and its local representative are
 \begin{equation}
 \begin{aligned}
\b\Omega= \begin{pmatrix} \Omega & \tfrac{1}{\ell}\Theta \\ \tfrac{-\epsilon}{\ell} \Theta^t & 0 \end{pmatrix}
\quad \text{ and } \quad 
\b F= \begin{pmatrix} F & \tfrac{1}{\ell} T \\ \tfrac{-\epsilon}{\ell} T^t & 0 \end{pmatrix} 
= \begin{pmatrix} R - \tfrac{1}{\ell^2}e\w e^T & \tfrac{1}{\ell} (de + A e) \\ \tfrac{-\epsilon}{\ell}(de + A e)^t & 0 \end{pmatrix}.
 \end{aligned}
 \end{equation} 
The normality condition is again $\Theta=0$, and the normal Cartan connection is s.t. $\omega=\omega(\theta)$ is the Levi-Civita connection. 
 The gauge transformations of the local Cartan connection and its curvature are
 \begin{equation}
 \begin{aligned}
{\b A}^\gamma&=\gamma\- \b A \gamma + \gamma\-d\gamma 
   &\quad \text{ and } \qquad
{\b F}^\gamma&=\gamma\- \b F \gamma
\\
&=\begin{pmatrix} \sS\- A \sS + \sS\-d\sS\  & \sS\- e \\
* & 0 \end{pmatrix}
& \quad
&=\begin{pmatrix} \sS\- F \sS & \sS\-T \\ * & 0 \end{pmatrix}.
 \end{aligned}
 \end{equation}
With linearisation, for the connection, still  $\delta_\chi \b A=\b D\chi$, with $\chi \in$ Lie$\SO(1,3)$, splitting as $\delta_\chi A =D\chi$ and $\delta_\chi e = -\chi e$. 
\medskip

\noindent \underline{{\bf Physics}}: The above Cartan-de Sitter geometry provides the kinematics for a gauge formulation of (vacuum) GR known as the MacDowell-Mansouri formulation \cite{McDowell-Mansouri1977, Stelle-West1979, Wise10}. 
Like YM theory, its Lagrangian is quadratic in the ($\so(1,3)$-part of the) curvature:
\begin{equation}
\begin{aligned}
L_{\text{{\tiny MM}}}(\b A) 
&=\tfrac{1}{2}\, F \eta\- \bullet F \eta\-
=\tfrac{1}{2}\, R \eta\- \bullet R \eta\- - \tfrac{\epsilon}{\ell^2} L_{\text{{\tiny EC}}}(\b A)\\
&= \tfrac{1}{2}\, F^{ab}F^{cd}\epsilon_{abcd} 
= \tfrac{1}{2}\, R^{ab} R^{cd}\epsilon_{abcd} - \tfrac{\epsilon}{\ell^2} \left(R^{ab} e^ce^d - \tfrac{\epsilon}{2\ell^2} e^ae^be^ce^d ... \right)\epsilon_{abcd}.
\end{aligned}
\end{equation}
The term quadratic in the Riemann 2-form $R$ is the Euler density of $M$; as such, it is called a topological term, and since it is a boundary term it does not change the field equation, but only the symplectic potential. The variational principle indeed gives: 
\begin{equation}
\begin{aligned}
\delta L_{\text{{\tiny MM}}}  &= \bs E_{\text{{\tiny MM}}}+d \bs\theta_{\text{{\tiny MM}}} = E(\delta \b A; \b A)+ d\theta(\delta \b A; \b A) \\
&= - \tfrac{\epsilon}{\ell^2}\left(\delta e \w e^T \bullet F + \delta A \bullet T \bullet e^T \right)
\ + \ d \left(\delta A \bullet F \right).
\end{aligned}
\end{equation}
The Einstein equation are still the piece of $\bs E_{\text{{\tiny MM}}}=0$ linear in $\delta e$, while the piece linear in $\delta A$ is the torsion-free condition.  The normal Cartan connection is again solution of the vacuum field equation. 

Compared with $L_{\text{{\tiny EC}}}$, the Euler density contributes $R\bullet e \w e^T$ to the potential $\bs\theta_{\text{{\tiny MM}}}$, from which one gets the $\SO(1,3)$-Noether charge 
\begin{equation}
\begin{aligned}
Q_\Sigma^{\text{{\tiny MM}}}(\chi; \b A)
= \int_\Sigma D\chi  \bullet F
\ \hat{=}\  \int_{\d \Sigma} \chi  \bullet F.
\end{aligned}
\end{equation}
The last on-shell equality uses only $T=0$.
We observe that the groundstate of the theory is again $(A)dS$, which is also the homogeneous model of the geometry, for which the Noether charge vanishes ($\b F=0$ implies $F=0$). A physically sensible feature. See \cite{Francois2021} for more comments.  
\smallskip

As for the Cartan-Riemann case, spinors belong naturally to this geometry as sections of associated bundles $\S= P \times_{\t\rho} \CC^{2n}$: Weyl $\CC^2$-spinors are associated to the spin represention $\t\rho: S\!O(1,3) \rarrow \text{spin}(1,3)=S\!L(2, \CC)$, Dirac $\CC^4$-spinors to $\t\rho': S\!O(1,3) \rarrow S\!L(2, \CC) \times S\!L(2, \CC)^*$ -- two Weyl spinors in conjugate representations.  The full theory coupled with matter fields, described by Dirac spinors $\psi$, is $L_{\text{{\tiny Grav}}}(\b A, \psi)=L_{\text{{\tiny MM}}}(\b A)+ L_{\text{{\tiny Dirac}}}(\b A, \psi)$. See e.g. \cite{Francois-et-al2021}.  
Via $\delta L_{\text{{\tiny Grav}}}(\b A)$ one obtains the field equations, 
$\bs E_{\text{{\tiny Grav}}}= \tfrac{-2\epsilon}{\ell^2}\delta e \big(e^T \bullet F - T(\psi, e)\big) + \tfrac{-2\epsilon}{\ell^2} \delta A \big(\bullet T\w e^T - S(\psi, e) \big) + \langle\delta \psi , \slashed D\psi - \mu \psi \rangle$, 
where $T(\psi, e)$ is the energy-momentum tensor of the Dirac field $\psi$ source of the Einstein equation, and $S(\psi, e)$ is the spin density of $\psi$ sourcing the torsion -- In the full theory, the Cartan connection cannot be normal. The Dirac equation $\slashed D\psi=\mu\psi$ describes the propagation of the Dirac field into $M$ as it interacts with the gravitational potential $\b A$; via an Eikonal or WKB (Wentzel-Kramers-Brillouin) approximation, one would relate the Dirac equation to the geodesic equations for a particle with non-zero spin/angular momentum (the Mathisson-Papapetrou-Pirani-Dixon equations) \cite{Rudiger1981}.

On-shell, one finds $Q_\Sigma^{\text{{\tiny Grav}}}(\chi; \b A, \psi)=Q_\Sigma^{\text{{\tiny MM}}}(\chi; \b A)$, meaning that the contribution of matter fields to the charge is measured by its sourcing of the gravitational fields (this is a gravitational Gauss law). 

\bigskip

\noindent $\bullet$ {\bf Conformal Cartan  geometry}:
This geometry $(P, \varpi)$ is based on the conformal group $G=S\!O(r+1,s+1)$  and  $H= C\!O(r,s) \ltimes  \RR^{m*}$, where $\RR^{4*}$ are special conformal transformations \cite{Ogiue, Kobayashi1972, Sharpe, Cap-Slovak09}. 
The homogeneous models is the conformally compactified Minkowski space, $G/H\simeq \b{\sf M}$. 
It is a $|1|$-parabolic geometry, the Lie algebra of $G$ splitting as $\LieG =\LieG_{-1}\oplus \LieG_0 \oplus\LieG_1= \RR^m \oplus \co(r,s) \oplus \RR^{m*}$, and the structure group being $\LieH=\LieG_0 \oplus\LieG_1=\co(r,s) \oplus \RR^{m*}$. 

In matrix notation, we write the gauge group 
$\Aut_v(P)\simeq \H =\H_0 \ltimes \H_1 = 
\begin{psmallmatrix} z & 0 & 0 \\
                     0 & \ \sS^{\phantom 0} & 0 \\
                     0 & 0 & z\-
\end{psmallmatrix}
\begin{psmallmatrix} 1 & r & \sfrac{1}{2}rr^t \\
                     0 & \1 & r^t \\
                     0 & 0 & 1^{\phantom 0}
\end{psmallmatrix}$, 
where $z\in C^\infty(M)$ is a Weyl rescaling function. 
The Cartan connection and its curvature are $\varpi= \varpi_{-1}+ \varpi_0 + \varpi_1 = \theta +\omega$ and 
$\b\Omega = \b\Omega_{-1}+ \b\Omega_0 + \b\Omega_1 = \Theta +\Omega$.
In matrix notation, their local representatives are
 \begin{equation}
 \begin{aligned}
\b A= \begin{pmatrix} 
a & K & 0 \\
e & A & K^t \\
0 & e^t & -a
\end{pmatrix}
\quad \text{ and } \quad 
\b F= 
\begin{pmatrix} 
-f & C & 0 \\
T & W & C^t \\
0 & T^t & -f
\end{pmatrix}.
 \end{aligned}
 \end{equation}
Often it is argued that, without loss of generality, the Weyl potential can be set to zero: $a=0$ (understood as a choice of gauge in physics,\footnote{Or as a so-called ``dressing operation". See  \cite{Attard-Francois2016_I, Attard-Francois2016_II}.}
as a choice of Weyl structure in mathematics \cite{Cap-Slovak09, Francois2019}). 

The soldering form induces a metric in the usual way, and a Hodge operator. 
Given that the gauge group acts on $\b A$ s.t. $e^\gamma=z\sS\-e$, it is clear that the induced metric $g=\eta \circ e$ gauge transforms as $g^\gamma=z^2g$. In other words, a conformal Cartan connection induces a conformal metric $[g]$ on $M$. 

For the (local) normal Cartan connection $\b A_{\text{{\tiny N}}}=\b A_{\text{{\tiny N}}}(e)$ the curvature is $\b F_{\text{{\tiny N}}}=\begin{psmallmatrix} 
0 & C & 0 \\
0 & W & C^t \\
0 & 0 & 0
\end{psmallmatrix}$ and s.t. $Ric(W)\defeq W^a_{bac}=0$, i.e. $W$ is the Weyl tensor. The $T=0$ condition allows to write $A=A(e)$. The condition $f=0$ enforces the symmetry of $K$ and $Ric(W)=0$ allows to write $K=K(A)$: 
This makes $K$ the Schouten tensor. So $C$, the covariant derivative of $K$, becomes the Cotton tensor. 
We see that the normality conditions allow to write $\b A_{i+1}$ in terms of $\b A_{i}$, so that in the end the nomal connection only depends on its soldering part $\b A_{-1}=e$. Clearly then, a normal conformal geometry $(P, \varpi_{\text{{\tiny N}}})$ encodes exactly the (gauge) geometry of a conformal manifold $(M, [g])$. 
\medskip

\noindent \underline{{\bf Physics}}: This geometry supplies the kinematics of a gauge model of gravity, once considered an alternative to GR, known as 4D ($m=4$) conformal gravity. To comply with the gauge principle it should be $\H$-invariant, \emph{not} $\SO(2,4)$-invariant. A Lagrangian quadratic in $\b F$ is a natural candidate. The Hodge dual of a (invariant) $p$-form $B$ gauge transforms as $(*B)^\gamma = z^{m-2p}(*B)$, in dim $m=4$ we have that $*\b F$ has the same tensoriality as $\b F$.\footnote{Notice it also means that YM theory is Weyl-invariant in dimension 4.} So the Lagrangian of 4D conformal gravity is
\begin{equation}
\begin{aligned}
L_{\text{{\tiny Conf}}}(\b A)=\Tr\big(\b F *\!\b F \big),
\end{aligned}
\end{equation}
in exact analogy with YM theory. 
The variational principle gives, similarly, 
\begin{equation}
\begin{aligned}
\delta L_\text{conf} &= \bs E_{\text{{\tiny conf}}}+d \bs\theta_{\text{{conf}}} = E(\delta \b A; \b A)+ d\theta(\delta \b A; \b A) \\
&= \Tr(\delta \b A\ \b D\!*\!\b F ) + d\Tr(\delta \b A *\!\b F), 
\end{aligned}
\end{equation}
which yields the YM equation for the conformal Cartan connection, $\b D*\!\b F=0$. 
In the normal case this reduces to 
\begin{equation}
\begin{aligned}
L_{\text{{\tiny Conf}}}(\b A_{\text{{\tiny N}}})=\Tr\big(\b F_{\text{{\tiny N}}} *\!\b F_{\text{{\tiny N}}} \big) = \Tr(W *\!W)=L_{\text{{\tiny Weyl}}}(e).
\end{aligned}
\end{equation}
The Lagrangian quadratic in the Weyl tensor $W$ defines Weyl gravity \cite{AFL2016_I}. It is known that the field equation for the latter, obtained by variation w.r.t. the soldering $e$ (or the metric $g$), is the Bach equation $B_{ab}=0$, with $B_{ab}$ the Bach tensor. 
Given the above, we have the immediate result that the Bach equation is encoded as the YM equation for the normal conformal Cartan connection:
\begin{equation}
\begin{aligned}
\b D_{\text{{\tiny N}}} *\!\b F_{\text{{\tiny N}}}=0 \quad \Leftrightarrow \quad B_{ab}=0.
\end{aligned}
\end{equation}
This  was first shown in a computational way in \cite{Korz-Lewand-2003}. Cartan geometry provides a straightforward conceptual proof. 

The spin version of conformal Cartan geometry in $m=4$ is based on the group $\b G=\text{Spin}(2,4)=S\!U(2,2)$, with the complexified Cartan connection 
$\b A^{\text{{\tiny $\CC$}}}$ and curvature
$\b F^{\text{{\tiny $\CC$}}}$ taking values in the $|1|$-parabolic Lie algebra $\b\LieG$.
The quadratic Lagrangian $L_{\text{{\tiny Conf}}}(\b A^{\text{{\tiny $\CC$}}})=\Tr\big(\b F^{\text{{\tiny $\CC$}}} *\!\b F^{\text{{\tiny $\CC$}}} \big)$ is that of  generalised twistor gravity. Indeed,  in the normal case this reduces to 
\begin{equation}
\begin{aligned}
L_{\text{{\tiny Conf}}}(\b A^{\text{{\tiny $\CC$}}}_{\text{{\tiny N}}})=\Tr\big(\b F^{\text{{\tiny $\CC$}}}_{\text{{\tiny N}}} *\!\b F^{\text{{\tiny $\CC$}}}_{\text{{\tiny N}}} \big) = \Tr(W^{\text{{\tiny $\CC$}}} *\!W^{\text{{\tiny $\CC$}}})=L_{\text{{\tiny Weyl}}}(\b e),
\end{aligned}
\end{equation}
where $W^{\text{{\tiny $\CC$}}}$ is the complexified Weyl tensor, $\b e \in $ Herm$(2, \CC)$ the complexified soldering form, and $\b A^{\text{{\tiny $\CC$}}}_{\text{{\tiny N}}}$ known as the twistor 1-form, in the sense of Penrose, as first noticed in \cite{Friedrich77}.
It thus follows that the YM equation for the twistor equation, the field equation for twistor gravity, encodes the Bach equation:
\begin{equation}
\begin{aligned}
\b D^{\text{{\tiny $\CC$}}}_{\text{{\tiny N}}} *\!\b F^{\text{{\tiny $\CC$}}}_{\text{{\tiny N}}}=0 \quad \Leftrightarrow \quad B_{ab}=0,
\end{aligned}
\end{equation}
where $\b D^{\text{{\tiny $\CC$}}}_{\text{{\tiny N}}}$ is known in physics as the ``twistor transport", or twistor connection \cite{Penrose-Rindler-vol1, Penrose-Rindler-vol2}. 
This was first shown, again computationaly, in \cite{Merkulov1984_I, Merkulov1984_II}. One may appreciate how Cartan geometry streamlines the proof. 

We invite the reader to consult \cite{Attard-Francois2016_II, Attard-Francois2016_I} for more details on the above and for the treatment of conformal tractors fields and twistor spinors in conformal Cartan geometry. 
We now turn to our description of the group manifold approach to gravity and supergravity.

\section{Supergroup manifold approach and Cartan supergeometry}
\label{Supergroup manifold approach and Cartan supergeometry}

In this section we will review the key aspects of the so-called (super)group manifold approach to (super)gravity, clarifying the physical and mathematical dictionary and making contact with Cartan supergeometry.

\subsection{Group manifold approach to pure gravity and Cartan geometry}
\label{Group manifold approach to pure gravity and Cartan geometry}

In 1978, Y. Ne'eman and T. Regge, driven by the growing interest on graded Lie algebras in physics, proposed a geometric approach to supersymmetric theories (in particular, supergravity). Such theories (and any gauge theory involving the action of a gauge group with nontrivial action on spacetime) are naturally formulated by using the concept of Grassmann algebras of forms in the context of the theory of Lie groups. As reported by the authors themselves \cite{Neeman-Regge1978b}, much of the necessary foundational work had been anticipated by Cartan and was already commonplace in mathematical literature at that time. However, it was not yet widely spread among physicists. Actually, even today, when within the physics community reference is made to the Cartan formulation of some physical theory, typically of gravity, the first (and, sometimes, unique) thing that come to mind is the use of differential forms, while, in fact, the differential (bundle) geometric setup adopted is much richer and constitutes a mathematical framework that lends itself particularly well to applications, extensions and generalizations.

The idea of Ne'eman and Regge, whose key aspects will be briefly reviewed in the following, consists of developing the formalism of gravitational theories as gauge theories on a \emph{(super)group manifold}, that is adding a differential  structure 
to the (super)group.\footnote{With the term supergroup we mean graded Lie groups.} Giving a manifold structure to a Lie (super)group $G$ allows to take limits and derivatives in $G$ and define the notion of a tangent space at the identity of $G$. The latter gives us the Lie (super)algebra $\mathfrak{g}$ associated with $G$, which is often easier to work with. All of this is standard in Klein geometry, where the use of the term ``group manifold'' is implicit and a principal $H$-bundle geometry for $G$ is assumed as a starting point. 

Klein geometry can be seen as the \emph{rigid} (or \emph{flat}) \emph{limit} (giving the \emph{vacuum configuration} in physics) of Cartan geometry, in which the Cartan curvature vanishes and the Cartan connection $\varpi$ boils down to $\varpi_G$ satisfying the Maurer-Cartan equations. Performing the rigid limit of a Cartan geometry $(P,\varpi)$ therefore amounts to considering the flat Cartan geometry $(P,\varpi)_\text{flat}$ locally isomorphic to the Klein geometry $(G,\varpi_G)$. 
On the other hand, one could say that a Cartan geometry is obtained by ``softening'' a Klein geometry. In the group manifold approach literature, this procedure, in general terms, is referred to as deforming the group manifold $G$ into a ``\emph{soft group manifold}'' $\tilde{G}$, the Maurer Cartan (MC) 1-form on $G$ being deformed into a 1-form on $\tilde{G}$ enjoying curvature. 
Let us describe the procedure as it appears in the physics literature. 

\subsubsection{Soft group manifold}
\label{Soft group manifold}

A soft group manifold $\tilde{G}$ is  endowed with a 
$\LieG$-valued 
1-form $\varrho^A$\footnote{Here, we use the abstract index convention:  $A$ is an abstract $\LieG$ index, no basis of the $\LieG$ is assumed.} 
with curvature
\begin{align}
    R^A := d \varrho^A + \tfrac{1}{2} {C^A}_{BC} \varrho^B \wedge \varrho^C \,.
\end{align}
where ${C^A}_{BC}$ are the structure constants of the Lie algebra $\mathfrak{g}$. The curvature fulfills the Bianchi identity $\nabla R^A=0$, where $\nabla$ is the ``$\tilde{G}$-covariant derivative",
w.r.t. $\varrho^A$.\footnote{To be understood in terms of the tractor derivative mentioned in section \ref{Cartan geometry in a nutshell}.}

Now, let $H \subset G$ be a subgroup of $G$ and $G/H$ the corresponding homogeneous manifold. The soft group manifold $\tilde{G}$ can be then considered as a principal bundle with fiber $H$ and base space $M$. Let $U_\alpha$ be a covering of $M$ and $\pi$ the projection $G \xrightarrow{{\small \pi}} M$; $\pi^{-1}(U_\alpha)\simeq U_\alpha \times H$ is parameterized by the elements $(x,y_\alpha)$, where $x\in M$ and $y_\alpha \in H$ -- a.k.a. local trivialisation of $\t G$.
We give transition functions $\varphi_{\alpha \beta}(x)\in H$ on $U_\alpha \cap U_\beta$ s.t., on $U_\alpha \cap U_\beta \cap U_\gamma$, $\varphi_{\alpha \beta} \varphi_{\beta \gamma}=\varphi_{\alpha \gamma}$. We identify the elements $(x,y_\alpha)=(x,y_\beta)$, where $y_\alpha = \varphi_{\alpha \beta} y_\beta$ and $x\in U_\alpha \cap U_\beta$. All the sets $\pi^{-1}(U_\alpha)$ are ``glued'' together into $\tilde{G}$. 
Over each $U_\alpha$, we have a map \begin{align}
 \chi_\alpha:    \pi^{-1} (U_\alpha) \subset \tilde{G} &\rarrow H \subset G \,, \\
   (x,y_\alpha) &\mapsto y_\alpha \,.
\end{align}
Given the left-invariant Maurer-Cartan form $\sigma^A$ on $G$, one has the form $\chi^\ast_\alpha \sigma^A$ on $\pi^{-1}(U_\alpha)\subset \tilde{G}$ with vanishing curvature. 
Then, given a $\mathfrak{g}$-valued 1-form $\tau^A_\alpha$ on $U_\alpha \subset M$, one requires that the restriction of $\varrho^A$ on $\pi^{-1}(U_\alpha)$
is s.t.
\begin{align}\label{fact}
    {\varrho^A}_\alpha = \chi^\ast_\alpha \sigma^A + \text{Ad}{(y_\alpha)_B}^A {\tau^B}_\alpha \,, \quad 
\end{align}
For this to hold for any $U_\alpha$ in the cover, 
the following matching conditions must hold over $U_\alpha \cap U_\beta$:
\begin{align}\label{matchcond}
    {\tau^A}_\beta = \varphi^\ast_{\alpha \beta} \,\chi^\ast_\alpha \sigma^A + \text{Ad}{(\varphi_{\alpha \beta})_B}^A {\tau^B}_\alpha  \,.
\end{align}
In the group manifold approach literature, if \eqref{fact}-\eqref{matchcond} hold, $\varrho^A$ is called a \emph{factorized} $c$-bein on the pair $(\tilde{G},H)$.

 As a matter of fact, $\varrho^A$ is indeed a Cartan connection on the $H$-bundle $\t G$. The name \emph{c-bein} alludes to the \emph{vielbein} of GR, or tetrad/(co)frame field, which provides a parallelism on spacetime $M$: it is meant to say that, likewise,  $\varrho^A$ provides a frame on the soft group manifold $\t G$. Which is exactly the distinctive defining property (iii) of a Cartan connection as define in section \ref{Cartan geometry in a nutshell}. The local representatives $\tau^A_\alpha$ on $U_\alpha \subset M$ are called ``Cartan gauges"  in \cite{Sharpe}, with  \eqref{matchcond} being their gluing relations. These are to be understood as the gravitational gauge potentials, with \eqref{matchcond}  representing their (passive) gauge transformations.
 

All of this applies when $G=\text{P}$ is the Poincaré group and $H=S\!O(1,3)$, with homogeneous flat model $G/H=\mathbb{R}^4$. Then, $\text{P}$ is softened to $\tilde{G}=\tilde{\text{P}}$ over spacetime $M$ locally modeled on the flat model. This is but the Cartan-Riemann geometry described in \ref{Special cases and application to physics}.  
But the idea can be extended, with adjustments, to the case of 
Lie supergroups, 
e.g. $G=\text{GP}$ the super-Poincaré group, softened to $\tilde{G}=\tilde{\text{GP}}$), a $H=S\!O(1,3)$-bundle over  the ``superspace" $M$ modeled on the super-Minkowski space $\mathbb{R}^{4|4}$ \cite{Neeman-Regge1978,Neeman-Regge1978b}. The first case is relevant to ``bosonic" pure gravity (vacuum GR),  as shown in \ref{Special cases and application to physics}. The second, is the arena for what is known as $\mathcal N =1$ pure supergravity (in four spacetime dimensions). 

Let us also mention that, at the algebraic level, there is a vector subspace $\mathfrak{f}$ of $\mathfrak{g}$ such that $\mathfrak{g}=\mathfrak{h}\oplus \mathfrak{f}$, $\mathfrak{f} \cap \mathfrak{h}=\emptyset$, with $\mathfrak{h}$ the Lie algebras of  $H$.
If $\mathfrak{f}$ is s.t. $\text{Ad}(h)\mathfrak{f}\subset \mathfrak{f}$, $h \in H$, $G/H$ is said to be a \emph{weakly reductive homogeneous manifold}. 
The $\LieG$-valued forms $\varrho^A$ and $R^A$ split accordingly.
This corresponds to the \emph{reductive Cartan geometries} described in \ref{Special cases and application to physics}.
In the case of pure bosonic gravity, for which $\tilde{G}=\tilde{\text{P}}$ and  $H=S\!O(1,3)$, the splitting gives $\varrho^A \rightarrow \lbrace \varrho^{ab},\varrho^a \rbrace$, $R^{A} \rightarrow \lbrace R^{ab},R^a \rbrace$ on $\t P$, and $\tau^A \rightarrow \lbrace \omega^{ab},V^a \rbrace$ on $U\subset M$ ( the indices  $a,b,\ldots \in {0,1,2,3}$ being for the fundamental vector representation of the Lorentz group). The local objects $\omega^{ab}$ and $V^a$ are commonly known in the physics literature as the ``\emph{spin connection}'' (in the general reductive case it is the $\mathfrak{h}$-components of the local representative of the Cartan connection) and the \emph{vielbein}  1-form $V^a={V^a}_\mu\, dx^{\,\mu}$, $\mu \in\{0, \ldots,3\}$, which is the \emph{local soldering form}).
The $\mathfrak{h}$-components of the curvature $R^A$ is called the \emph{proper curvature} of $\varrho^A$  (the $\mathfrak{h}$-components of the \emph{Cartan curvature}), while the $\mathfrak{f}$-components represent the \emph{torsion of the Cartan connection}. In terms of their \emph{local representatives}, we have, respectively, the Riemann curvature 2-form and the torsion 2-form
\begin{align}
    \mathcal{R}^{ab} & = d \omega^{ab} + {\omega^a}_c \wedge \omega^{cb} \,, \\
    \mathcal{R}^a & = dV^a + {\omega^a}_b \wedge V^b = D V^a \,,
\end{align}
where $D$ denotes the Lorentz-covariant derivative. Compare with   \eqref{Cartan-co-CR}-\eqref{Cartan-curv-CR} in \ref{Special cases and application to physics}. 
Table \ref{Dico2} below  gives summary of the dictionary between the notations used above, typical of the group manifold approach endowed with a principal bundle structure, and the ones used in section \ref{Classical gauge theory of gravity and Cartan geometry} to describe  Cartan geometry.
\medskip
\begin{table}[!h]
\begin{center}
\begin{tabular}{||c | c||}
 \hline
  & \\
 Group manifold approach & Cartan geometry \\
 [2ex] 
 \hline\hline 
  & \\
 $\tilde{G}$; \,\, $H$; \,\, $M$  & $P$; \,\, $H$; \,\, $M$    \\ [1.5ex] 
 $\varrho$ \,\, (factorized $c$-bein, Cartan connection) & $\varpi$ \,\, (Cartan connection) \\[1.5ex]
 $\tau$ \,\, (local representative of $\varrho$) & $\bar{A}$ \,\, (local representative of $\varpi$) \\[1.5ex]
 $\omega^{ab}$ \,\, (``spin connection'') & $A^{ab}$ \,\, ($\mathfrak{h}$-components of the local rpr. of $\varpi$)  \\[1.5ex]
 $V^{a}={V^a}_\mu dx^\mu$ \,\, (``vielbein'', meaning local soldering form) & $e^{a}={e^a}_\mu dx^\mu$ \,\, (local soldering form) \\[1.5ex]
 $R=\lbrace R^{ab},R^a\rbrace$ \,\, (Cartan curvature, or proper curvature) & $\bar{\Omega}=\lbrace \Omega^{ab}, \Theta^a \rbrace$ \,\, (Cartan curvature) \\[1.5ex]
 $\mathcal{R}^{ab}$ \,\, (Riemann curvature 2-form) & $R^{ab}$ \,\, (Riemann curvature 2-form) \\[1.5ex]
 $\mathcal{R}^{a}$ \,\, (torsion 2-form, local rep. of the torsion $R^{a}$ of $\varrho$) & $T^{a}$ \,\, (torsion 2-form, local rep. of the torsion $\Theta^{a}$ of $\varpi$) \\
\phantom{basta!} & \\
\hline
 \end{tabular}
 \caption{\label{Dico2}Group manifold approach and Cartan geometry notations.}
 \end{center}
\end{table}

\subsubsection{Geometric Lagrangian and action for gravity}
\label{Geometric Lagrangian and action for gravity}

It is not our aim here to  review in details the group manifold approach to pure gravity, given that it ends up being equivalent to the Cartan formulation already discussed.\footnote{For concise reviews of the subject we refer the reader to, e.g., \cite{DAuria:2020guc,Castellani:2019pvh}.} 
However, we  sketch how  the  action for pure GR is constructed in this approach, as it provides the template for the  supersymmetric case considered next.
One considers the following Lagrangian 
on $\tilde{\text{P}}$:\footnote{The building rule for a ``geometric'' Lagrangian adopted in the (super)group manifold approach can be found in \cite{Castellani:1991eu}.}
\begin{align}
    \mathcal{L} = R^{ab} \wedge \varrho^c \wedge \varrho^d \epsilon_{abcd} \,.
\end{align}
As it is a form on $\t P$, 
to obtain the action we need to 
integrate over a 4-dimensional submanifold. One thus considers a smooth  embedding map $\sigma: M^4 \rightarrow \tilde{\text{P}}$, so that:
\begin{align}\label{gractgm}
    \mathcal{S} = \int_{\sigma(M^4)} \mathcal{L} = \int_{M^4} \sigma^\ast \mathcal{L} = \int_{M^4} \mathcal{R}^{ab} \wedge V^c \wedge V^d \epsilon_{abcd} \,,
\end{align}
The last integral in \eqref{gractgm} is performed over $M^4$, identified with spacetime. So, the action defined as the integral of $\mathcal{L}$ over a section of $\tilde{\text{P}}$ is equivalent to the Einstein-Cartan action for pure gravity, based on Cartan-Riemann geometry. 
We observe that the above embedding should be understood as a \emph{global} section of the principal bundle $\t P$, which is then a trivial bundle over spacetime $M^4$: $\t P= M^4 \times H$. 

One usually argue for, or requires, independence of the integration performed in \eqref{gractgm} from the choice the section $\sigma$.
This amounts to  requiring that $\L$ be \emph{basic} on $\t P$, i.e. that $\s^* \L$ be invariant under gluings on $M$:  in other words, it amounts to requiring Lorentz gauge invariance of $\s^* \L$. 
Which is the gauge principle requirement.
\footnote{We remark that basicity of $\L$ implies in particular that the curvature $R^{ab}$ is \emph{horizontal}. So, the Lie derivative of the Cartan connection $\varrho^A$ on $\tilde{\text{P}}$ along vertical vector fields is an infinitesimal gauge transformation. Let us also observe that what are typically called  ``general coordinate transformations" (GCTG) in the group manifold approach literature (together with their anholonomized version; see e.g. \cite{Neeman-Regge1978b,DAuria:2020guc}) are just diffeomorphisms in $\tilde{\text{P}}$, whose infinitesimal version is just given by the action of the Lie derivative along the generating vector field.}

The field equations obtained by varying the action \eqref{gractgm} w.r.t. $\omega^{ab}$ and $V^a$ read, respectively,
\begin{align}
    \mathcal{R}^c \wedge V^d \epsilon_{abcd} = 0 \quad & \Rightarrow \quad \mathcal{R}^a=0 \,, \, \text{as, setting $\mathcal{R}^a={\mathcal{R}^a}_{\mu \nu}dx^\mu dx^\nu$, we get ${\mathcal{R}^a}_{\mu \nu}=0$} \,, \\
    \mathcal{R}^{ab} \wedge V^c \epsilon_{abcd} = 0 \quad & \Rightarrow \quad \text{vacuum Einstein's equations.}
\end{align}
Let us also mention that one may try to extend the integration of $\mathcal{L}$ to the entire $\tilde{P}$. In this case, one should write an action
\begin{align}
    \mathcal{S} = \int_{\tilde{P}}  R^{ab} \wedge \varrho^c \wedge \varrho^d \epsilon_{abcd} \wedge \nu \,,
\end{align}
where $\nu$ is a 6-form including the $dy$ differentials. As far as we know, no explicit form for $\nu$ has ever been proposed such that the theory reproduces Einstein's equations. However, a similar idea, based on the notion of Poincaré dual and exploiting integral and pseudo forms (and introducing so-called ``picture changing operators''), was considered in the supersymmetric case in more recent literature \cite{Castellani:2014goa,Castellani:2019pvh} as a way to perform integration of a superspace geometric Lagrangian over the entire superspace, rather than to a bosonic submanifold (identified with spacetime) immersed in superspace, as it is done in the supergroup manifold approach.

The implementation of the above setup to recover the MacDowell-Mansouri formulation \cite{McDowell-Mansouri1977, Stelle-West1979,Wise10} of GR, whose kinematics is provided by Cartan-de Sitter geometry, can be found in \cite{Neeman-Regge1978b}. The scheme can be applied to derive  $4D$ conformal gravity. 
A more complete list of building rules for a general geometric Lagrangian can be found in \cite{Castellani:1991eu,DAuria:2020guc}, together with several applications both at the ``purely bosonic'' level and to supergravity.

Usually, in the (super)gravity literature based on the (super)group manifold approach, a principal bundle structure with the Lorentz group as the structure group is assumed from the very beginning. On the other hand, supersymmetry transformations in supergravity are (usually) not gauge/vertical transformations of a bundle. So there is no simplifying factorization of the odd Grassmann direction. As we are going to discuss in the following, the extension of the above treatment to the case of supergroup manifolds  yields a \emph{geometric interpretation of supersymmetry}, in which supersymmetry transformations are odd diffeomorphisms  of superspace $M^{4|4}$.


\subsection{Supergroup manifold (a.k.a. ``rheonomic'') approach to SUGRA}
\label{Supergroup manifold approach (a.k.a. ``rheonomic'') to SUGRA}

The machinery of the (soft) group manifold approach, underpinned by Cartan geometry, 
has been to extended to supergroups: it is known as  the (soft) supergroup manifold approach, a.k.a. \emph{rheonomic approach} \cite{Castellani:1991eu}, and has been applied to build (or re-derive ``geometrically'') both supergravity and rigid supersymmetric theories. Here our focus will be on supergravity. We will see how it hints at  Cartan supergeometry.

A supergroup manifold $G$ (softened to $\tilde{G}$) has both even (commuting, or ``bosonic'') and odd (``fermionic'') coordinates: $(x^{\,\mu},y^{\,\mu\nu},\theta^{\,\alpha})$, where the $\theta$'s are the fermionic Grassmann coordinates and $\alpha,\beta,\ldots$ are spinor indices, which we will generally omit in the following. For simplicity, we will restrict ourselves to the $\mathcal{N}=1$ 
case in four spacetime dimensions. 
In $\mathcal{N}$-extended supersymmetric models, where $\mathcal{N}$ is the number of supercharges, the $\theta$'s also carry a so-called R-symmetry index $\mathcal{A}=1,\ldots,\mathcal{N}$. The framework can be generalized to higher spacetime dimensions too. 

A principal superbundle structure is considered for $\tilde{G}$, with $H=S\!O(1,3)$ (vertical Lorentz directions), and the superspace $M^{4|4}$ is modeled on the homogeneous super-Minkowski space $\mathbb{R}^{4|4}$, so has coordinates $(x^\mu,\theta^\alpha)$. In the pure supergravity case, and in the absence of cosmological constant, one considers  $\tilde{G}=\tilde{\text{GP}}$ (soft graded Poincaré).  
This setup corresponds to a Cartan supergeometric framework, where the local representative of the \emph{Cartan superconnection} $\rho^A$ on $\t{\text{GP}}$  is  a (set of) 1-form(s) \emph{superfield(s)} $\tau^A$ on $M^{4|4 }$, with \emph{supercurvatures} $\mathcal{R}^A$. 
The superfield $\tau^A$ contains the \emph{supervielbein} $\lbrace V^a ,\psi^\mathcal{\alpha}\rbrace$, which is an orthonormal basis of 1-forms at each point of the cotangent plane to  superspace $M^{4|4}$ ($V^a$ is commonly referred to as the ``bosonic vielbein'', while $\psi^\mathcal{\alpha}$ is the ``fermionic'' one, also called \emph{gravitino} 1-form). 
Besides them, $\tau^A$ also includes the Lorentz spin connection $\omega^{ab}$, as a 1-form superfield. 

The supercurvatures $\mathcal{R}^A$,
\begin{align}
    \mathcal{R}^A := d \tau^A + \tfrac{1}{2} {C^A}_{BC} \tau^B \wedge \tau^C \,, \quad 
\end{align}
with ${C^A}_{BC}$ the structure constants of the Lie  superalgebra $\LieG$ (super-Poincaré algebra), 
are horizontal w.r.t. the Lorentz directions. But, since $M^{4|4}$ does not decompose into a bundle over $M^4$ with Grassmanian fibers, $\mathcal{R}^A$ is not ``horizontal"  in the fermionic direction, along of the fermionic vielbein. This implies that local supersymmetry transformations are not gauge tranformations. They correspond instead to diffeomorphisms along the $\theta$-directions of superspace (given linearly by the Lie derivative $\ell_{\epsilon} := \iota_{\epsilon} d + d \iota_\epsilon$ along the generating $\epsilon^\alpha=\delta\theta^\alpha$ supersymmetry (odd) vector field).

The superfields $\tau^A(x,\theta)$, and their supercurvatures $\mathcal{R}^A(x,\theta)$ decomposes locally on superspace as 
\begin{align}
    \tau^A(x,\theta) & = {\tau^A}_\mu (x,\theta) dx^\mu + {\tau^A}_\alpha (x,\theta) d\theta^\alpha \,, \\
    \mathcal{R}^A(x,\theta) & = {\mathcal{R}^A}_{\mu \nu} (x,\theta) dx^\mu \wedge dx^\nu + {\mathcal{R}^A}_{\mu \alpha} (x,\theta) dx^\mu \wedge d\theta^\alpha + {\mathcal{R}^A}_{\alpha \beta} (x,\theta) d\theta^\alpha \wedge d\theta^\beta \nonumber \\
    & = {\mathcal{R}^A}_{\mu \nu} (x,\theta) dx^\mu \wedge dx^\nu + {\mathcal{R}^A}_{L \alpha} (x,\theta) dZ^L \wedge d\theta^\alpha \,,
\end{align}
where $dZ^L=(dx^\mu,d\theta^\alpha)$.
The superfunction ${\mathcal{R}^A}_{L \alpha}(x,\theta)$ are called the ``outer'' components of $\mathcal{R}^A(x,\theta)$, while ${\mathcal{R}^A}_{\mu \nu} (x,\theta)$ are called the ``inner" components (or \emph{supercovariant field strengths}). 
The superfield $\tau^A(x,\theta)$ induces the corresponding ``spacetime quantity'' ($x$-fields)  $\tau^A(x)={\tau^A}_\mu (x) dx^\mu$ on $M^4 \subset M^{4|4}$ via the
restriction
\begin{align}
    \tau^A(x) = \tau^A(x,\theta)|_{\theta=d\theta=0} = {\tau^A}_\mu(x,0) dx^\mu \,.
\end{align}
Note that each component in the $\theta$-expansion of a superfield represents, \emph{a priori}, a new $x$-space field. Hence, \emph{a priori}, the theory defined in superspace could exhibit extra dynamics (extra degrees of freedom) w.r.t. the spacetime restriction. If this were the case, the resultant theory would fail to be
equivalent to supergravity formulated in terms of a local spacetime (super)symmetry. As we are going to discuss, what allows the formulation of supergravity in
superspace to be equivalent to the one on spacetime is the so-called \emph{rheonomy principle}.

\medskip

\noindent \underline{{\bf Rheonomy}}: Let us consider the so-called \emph{rheonomic extension mapping}
\begin{align}\label{rhextmap}
    \tau^A(x) \rightarrow \tau^A(x,\theta) \,.
\end{align}
The knowledge of this mapping is crucial in order to interpret the theory based on the superspace fields as a spacetime theory. Indeed, in order to have the same physical content as the spacetime theory, we must be able to determine the fields contained in the $\theta$-expansion of $\tau^A(x,\theta)$, and all its $d\theta$ components, in terms of its spacetime restriction ${\tau^A_\mu}(x,0)dx^\mu$, that is what the knowledge of \eqref{rhextmap} amounts to. 

For the mapping to be fully determined, a complete set of Cauchy data must be known: not only ${\tau^A}_\mu(x,0)$ but also the ``normal derivatives'' $\left(\frac{\partial}{\partial \theta^\alpha} {\tau^A}_\mu(x,\theta)\right)|_{\theta=0}=\frac{\partial}{\partial \theta^\alpha} {\tau^A}_\mu(x,0)$. This, in general, is not the case. 
Now, considering the diffeomorphic mapping generated by the Lie derivative $\ell_\epsilon$, $\epsilon=\epsilon^\alpha \frac{\partial}{\partial \theta^\alpha}$, one can prove \cite{Castellani:1991eu} that the knowledge of $\tau^A(x,0)$ (that is, besides ${\tau^A}_\mu(x,0)$, also ${\tau^A}_{\alpha}(x,0)$) and $\left(\frac{\partial}{\partial \theta^\alpha} {\tau^A}(x,\theta)\right)|_{\theta=d\theta=0}=\frac{\partial}{\partial \theta^\alpha} {\tau^A}_\mu(x,0)dx^\mu$ is equivalent to the knowledge of ${\tau^A}_\mu(x,0)$ only, together with $\mathcal{R}^A_{\alpha \mu}(x,0)$. 
Actually, releasing the $d\theta=0$ restriction, one can also show that the knowledge of ${\tau^A}(x,0)$ and $\frac{\partial}{\partial \theta} {\tau^A}(x,0)$ is equivalent to that of ${\tau^A}_\mu(x,0)$ together with ${\mathcal{R}^A}_{L\alpha}(x,0)$.

The concept of rheonomy can now be introduced. Assume the following constraints (a.k.a. \emph{rheonomic constraints}) to hold:
\begin{align}\label{rhconstr}
    {\mathcal{R}^A}_{L\alpha} = {C^{A|\mu\nu}}_{L\alpha|B} {\mathcal{R}^B}_{\mu \nu} \,,
\end{align}
where ${C^{A|\mu\nu}}_{L\alpha|B}$ are suitable invariant tensors of the supergroup. Then, the knowledge of a purely spacetime configuration $\lbrace {\tau^A}_\mu(x,0),\partial_\mu {\tau^A}_\nu(x,0) \rbrace$ determines in a complete way the extension mapping \eqref{rhextmap}. \\
In other words, if \eqref{rhconstr} hold, given a purely spacetime configuration the complete $\theta$-dependence of the associated superfields $\tau^A(x,\theta)$ can be recovered, as ${\tau^A}_\mu(x,0)$ and $\partial_\mu {\tau^A}_\nu(x,0)$ (or, equivalently, ${\tau^A}_\mu(x,0)$ and ${\mathcal{R}^A}_{\mu \nu}(x,0)$) constitute a complete set of Cauchy data on spacetime when \eqref{rhconstr} is satisfied. \\
The property expressed by \eqref{rhconstr} was named ``rheonomy'' and a theory admitting a set of rheonomic constraints is likewise said to be rheonomic.\footnote{From the ancient Greek words ``rhein'', which means flow, and ``nomos'', which means law, referring to the lift from an $x$-space configuration $\tau^A(x,0)$ to a superspace one, $\tau^A(x,\theta)$ (``\emph{passive}'' interpretation of the Lie derivative along the odd directions of superspace).}
From a physical viewpoint, the restriction on the superspace parametrization of the supercurvatures given by the rheonomic constraints guarantees that no additional degree of freedom is introduced in the theory in superspace compared to those already present in spacetime.

Alternatively, if we regard the Lie derivative as generating the functional change of $\tau^A$ at the same coordinate point (``\emph{active}'' interpretation of the Lie derivative),
\begin{align}
    \ell_\epsilon \tau^A = {\mu^A}'(x,0)-\mu^A(x,0) \,,
\end{align}
it transforms a given spacetime configuration into a new spacetime configuration (\emph{supersymmetry transformations}). 

In particular, invariance of the theory, i.e. of the action (or the field equations), under diffeomorphisms along the odd directions in superspace amounts to supersymmetry invariance of the theory restricted to spacetime. One may write, schematically \cite{Castellani:1991eu},

\begin{equation}\label{sgmapprmap}
\begin{aligned}
\text{Superspace configuration $\tau^A(x,\theta)$} \quad & \xrightarrow{\,\,\text{$\epsilon^\alpha$-diffeomorphisms}\,\,} \quad \text{New superspace configuration ${\tau^A}'(x,\theta)$} \\
\uparrow{\text{\small{Rheonomy}}} \quad \quad \quad & \quad \quad \quad \quad \quad \quad \quad \quad \quad \quad \quad \quad \quad \quad \downarrow{\text{\small{Restriction to $x$-space}}} \\
\text{$x$-space configuration $\tau^A(x,0)$} \quad & \xrightarrow{\,\,\,\,\,\,\,\text{Supersymmetry}\,\,\,\,\,\,\,} \quad \text{New $x$-space configuration ${\tau^A}'(x,0)$}
\end{aligned}
\end{equation}

\medskip

Demanding the closure of the Lie derivative brackets (that is, requiring the induced transformations to form a Lie super-algebra) is then equivalent to demanding integrability of the rheonomic constraints, implying constraints on the inner components of the supercurvature.
Checking that $d^2=0$ amounts to checking that the Bianchi identities are satisfied by the supercurvatures $\mathcal{R}^A$.\footnote{The Bianchi identities guarantee the closure of the given algebra when represented in terms of fields.}
In the presence of \eqref{rhconstr}, the Bianchi identities loose the character of identities and become integrability equations for the constraints. 
Since the rheonomic constraints express the outer components ${\mathcal{R}^A}_{L\alpha}$ of $\mathcal{R}^A$ in terms of the inner ones ${\mathcal{R}^A}_{\mu \nu}$,  the Bianchi relations (Bianchi-integrability equations) are equations among the ${\mathcal{R}^A}_{\mu \nu}$'s which must be valid everywhere in superspace. In particular on the restriction to the spacetime hypersurface. 
Hence, we conclude that the supersymmetry transformations algebra close only if ${\mathcal{R}^A}_{\mu \nu}$ satisfy certain integrability equations, obtained from the Bianchi ``identities''. 

Physically, these equations are  the spacetime field equations of the theory;  we may thus say that the supersymmetry algebra closes \emph{on-shell}. 
In this respect, the supersymmetry algebra is peculiar since supersymmetry representations have to contain the same number of bosonic and fermionic degrees of freedom. However, the on-shell condition changes in different ways the number of degrees of freedom (d.o.f.) of fields of different spin (e.g. spinors halve their d.o.f., gauge vectors lower by one
their d.o.f., and scalars do not change them). As a consequence, when supersymmetry is realized in terms of field representations (supermultiplets), as it happens in supergravity
theories, it is an on-shell symmetry.\footnote{This restriction can be relaxed by including auxiliary fields in the theory, but, considering a finite number of auxiliary fields, this is possible only in few cases, while the most fruitful approach so far has been that of the so-called \emph{harmonic superspace} \cite{Galperin:2001seg} (where spherical harmonics are employed), involving an infinite number of auxiliary fields. In these cases, the Bianchi identities are proper identities and the supersymmetry algebra closes \emph{off-shell}; see, e.g., \cite{Ravera:2022vjy} for a concise discussion on this topic.}

Therefore, in a rheonomic theory (in the absence of auxiliary fields) we expect the supersymmetry transformations to close an algebra only on the on-shell configurations of $\tau^A(x,0)$, implying that we can lift to superspace only those configurations which are solutions of the $x$-space field equations, while arbitrary configurations cannot be lifted.

\subsubsection{Geometric Lagrangian in superspace and action}
\label{Geometric Lagrangian in superspace and action}

The on-shell closure of the Bianchi ``identities'' provides the field equations that one may  obtain from a supersymmetric Lagrangian  $\mathcal{L}$ in superspace: i.e. satisfying $\delta_\epsilon \mathcal{L}=\ell_\epsilon \mathcal{L}=0$, maybe up to boundary terms (note that $\mathcal{L}$ is not a top form in superspace).\footnote{We refer to, e.g., \cite{Castellani:1991eu,DAuria:2020guc,Andrianopoli:2014aqa,Concha:2018ywv} for a variety of simple models presented and studied in details. Furthermore, building rules for a ``geometric'' superspace Lagrangian can be found in, e.g., \cite{DAuria:2020guc}.
}
On the supergroup manifold $\tilde{G}=\tilde{\text{GP}}$, given a Lagrangian density $L^{(4|4)}$, one may write the following action:
\begin{align}
    \mathcal{S} = \int_{\sigma(M^{4|4})} L^{(4|4)} = \int_{M^{4|4}} \sigma^\ast L^{(4|4)} =\int_{M^{4|4}} \mathcal{L}^{(4|4)} \,, 
\end{align}
where $\sigma(M^{4,4})$ is a (global) section of $\tilde{\text{GP}}$. The latter being an $H=S\!O(1,3)$-principal superbundle, Lorentz invariance (basicity) of $L^{4|4}$ ensures freedom in the choice of the section of integration. On the r.h.s., $M^{4|4}$ is identified with  superspace. This prescription is along the lines of, e.g., \cite{Castellani:2014goa,Castellani:2019pvh}, where the idea is to construct a full superspace Lagrangian $\mathcal{L}^{4|4}=\mathcal{L}^{(4|0)}\wedge \mathcal{Y}^{(0|4)}$, where $\mathcal{Y}^{(0|4)}$ is a so-called ``\emph{picture changing operator}'', PCO, a differential object ``inherited'' from the string field theory literature. 
This methods allows integration over the entire $M^{4|4}$. PCOs are typically built by hand and different choices of PCO can lead to different realisations of the same theory. 
A particular choice of  $\mathcal{L}^{(4|0)}$ is the rheonomic 4-form Lagrangian, a bosonic 4-form in superspace:
\begin{align}
    \mathcal{L}^{(4|0)} = \mathcal{R}^{ab} \wedge V^c \wedge V^d \epsilon_{abcd} + 4 \bar{\psi} \wedge \gamma_5 \gamma_a D \psi \wedge V^a \,.
\end{align}
The first term is the the so-called Einstein-Hilbert, the second is a  Rarita-Schwinger term. It is built with the superfields $\lbrace{\omega^{ab},V^a,\psi\rbrace}$ and the $\gamma$'s are Dirac gamma matrices. \\
The supercurvatures (components of the local representative of the supercurvature 2-form of the Cartan superconnection) are
\begin{align}
    \mathcal{R}^{ab} & = d \omega^{ab} + {\omega^a}_c \wedge \omega^{cb} \,, \\
    \mathcal{R}^a & = dV^a + {\omega^a}_b \wedge V^b - \frac{i}{2}\bar \psi \wedge \gamma^a \psi = D V^a - \frac{i}{2}\bar \psi \wedge \gamma^a \psi \,, \\
    \rho & = d \psi + \frac{1}{4} \omega^{ab} \wedge \gamma_{ab} \psi = D \psi \,,
\end{align}
where $\mathcal{R}^a$ is called the supertorsion and $\rho$ is the gravitino super field strength. 
The field equations read\footnote{Up to boundary terms.}
\begin{align}
    & \mathcal{R}^c \wedge V^d \epsilon_{abcd} = 0 \,, \\
    & \mathcal{R}^{ab} \wedge V^c \epsilon_{abcd} - 2 \bar \psi \wedge \gamma_5 \gamma_d D \psi = 0 \,, \\
    & 2 \gamma_5 \gamma_a D \psi \wedge V^a - \gamma_5 \gamma_a \psi \wedge \mathcal{R}^a = 0 \,.
\end{align}
The analysis of their expansion along $VVV$, $VV\psi$, $V\psi\psi$, $\psi\psi\psi$ yields the spacetime field equations together with the rheonomic constraints on the outer components of the supercurvatures, which therefore results to be expressed, on-shell, as linear tensor combinations of the inner components.

On the other hand, in the original formulation of the geometric approach to supergravity in superspace, the rheonomic Lagrangian 
$\mathcal{L}^{(4)}=\mathcal{L}^{(4|0)}$ is integrated over a bosonic hypersurface, 
opportunely identified with spacetime $M^4$,\footnote{This is coherent, given the on-shell character of the construction. Indeed, once a geometric supersymmetric (typically, up to total divergences) Lagrangian is constructed, in the spirit of \cite{Castellani:1991eu}, the rheonomic constraints are recovered from the study of the field equations, meaning that the theory on another bosonic submanifold immersed in superspace, with a different embedding, would be given just by the supersymmetry transformed Lagrangian (plus a Lorentz gauge transformation, if we think of the  superspace $M^{4|4}$ in turn immersed in the  supergroup manifold), therefore leaving unaltered the physical content.} immersed in superspace. Let us mention that, in this setup, the invariance of the action does not coincide, in general, with the invariance (typically, up to total divergences) of the Lagrangian. In particular, assuming superspace to be compactified along the spacetime directions, the diffeomorphisms in the odd directions of superspace are an (off-shell) invariance of the action $\mathcal{S}$ built in this setup iff 
\begin{align}
   \mathcal{S}(M^4+\delta M^4)-\mathcal{S}(M^4) = \int_{\mathcal{V}} d\mathcal{L}^{(4)} = 0 \quad \rightarrow \quad d \mathcal{L}^{(4)} = 0 \quad \text{(i.e., if $\mathcal{L}^{(4)}$ is a closed form in superspace).}
\end{align}
We have denoted by $\mathcal{V}$ the supervolume contained between the two hypersurfaces ${M^4}'=M^4+\delta M^4$ and $M^4$.

\medskip

The application of the rheonomic approach to, e.g., the case $\tilde{G}=\tilde{O\S\!p(1|4)}$ and to the supersymmetric extension of the MacDowell-Mansouri formulation for gravity (``MacDowell-Mansouri supergravity'', we might say, based on Cartan-(Anti) de Sitter supergeometry) can be found, respectively, in \cite{Castellani:1991eu,DAuria:2020guc} and \cite{Andrianopoli:2014aqa,Andrianopoli:2021rdk}. The same construction can also be applied to the superconformal theory (see \cite{DAuria:2021dth}). A rather comprehensive list of applications and more recent results obtained in the rheonomic approach is to be found in \cite{Castellani:2019pvh,DAuria:2020guc}.\footnote{See also \cite{Castellani:1990nr} for a geometric interpretation of the BRST symmetry developed within the group manifold setup.}

\subsubsection{Higher forms supergravities and FDAs}
\label{Higher forms supergravities and FDAs}

Supergravity theories in $4 \leq D \leq 11$ spacetime dimensions have a bosonic field content that generically includes, besides the metric and a set of 1-form gauge potentials, also $p$-index antisymmetric tensors.\footnote{An example in the geometric supergravity literature is given by the 3-index antisymmetric tensor $A_{\mu\nu\rho}$ of the Cremmer-Julia-Scherk $D=11$ supergravity theory \cite{Cremmer-et-al1978}, which is described geometrically in terms of a 3-form gauge potential $A^{(3)}$ with super field strength $F^{(4)}=dA^{(3)}-\frac{1}{2} \bar\Psi \wedge \Gamma_{ab}\Psi \wedge V^a \wedge V^b$ \cite{DAuria:1982uck}.} Their vacuum structure are therefore appropriately discussed in the framework of so-called ``Free Differential Algebras'' (FDAs) underlying the theory, which extend the Maurer-Cartan equations by incorporating $p$-form gauge potentials. The concept of FDA was introduced by Sullivan in \cite{Sullivan:1977}. Subsequently, the FDA framework was applied to the study of supergravity theories by R. D'Auria and P. Fré, in particular in \cite{DAuria:1982uck}, where the FDA was referred to as \emph{Cartan Integrable System} (CIS), since the authors were unaware of the previous work by Sullivan. In fact, FDA and CIS are equivalent concepts \cite{DAuria:1982ada}. The latter is also known as the \emph{Chevalley-Eilenberg Lie algebras cohomology} (CE-cohomology) framework in supergravity. Actually, the super algebraic structures called FDAs in the geometric supergravity literature are \emph{super semifree differential graded-commutative algebras}.\footnote{Nowadays, one would relate them to super $L_\infty$-algebras.}

Let us briefly recall the standard procedure to construct a \emph{minimal} FDA (a minimal FDA is one where the differential of any $p$-form does not contain forms of degree greater than $p$), starting from an ordinary Lie algebra. Then, we will discuss the extension to the supersymmetric case. 

Let us thus start by considering the Maurer-Cartan 1-forms $\sigma^A$ of a Lie algebra, and let us construct the so called \emph{$(p+1)$-cochains} (\emph{Chevalley cochains}) $\Omega^{i|(p+1)}$ in some representation $D^i_j$ of the Lie group, that is to say, $(p+1)$-forms of the type
\begin{align}\label{coch}
   \Omega^{i|(p+1)}=\Omega^i_{A_1\dots A_{p+1}}\sigma^{A_1}\wedge \dots \wedge \sigma^{A_{p+1}} \,,
\end{align}
where $\Omega^i_{A_1\dots A_{p+1}}$ is a constant tensor. 
If the above cochains are closed, 
\begin{align}
d\Omega^{i|(p+1)}=0 \,,
\end{align}
they are \emph{cocycles}. If a cochain is exact, it is called a \emph{coboundary}. 
Of particular interest are those cocycles that are not coboundaries, which are elements of the CE-cohomology.\footnote{If the cocycles are also coboundaries, then the cohomology class is trivial.} 
In the case in which this happens, we can introduce a $p$-form $A^{i| (p)}$ and write the following closed equation:
\begin{align}\label{enlarge}
 d A^{i|(p)}+\Omega^{i| ({p+1})}=0 \,,
\end{align}
which, together with the Maurer-Cartan equations of the Lie algebra, is the first germ of a FDA. The latter contains, besides the Maurer-Cartan 1-forms $\sigma^A$, also the new $p$-form $A^{i|(p)}$. \\
The procedure can be iterated taking as basis of new cochains $\Omega^{j|(p'+1)}$ the full set of forms, namely $\sigma^A$ and $A^{i | (p)}$, and looking again for cocycles.  
If a new cocycle $\Omega^{j| (p'+1)}$ exists, then we can add to the FDA a new equation,
\begin{align}\label{enlarge1}
 d A^{j \vert (p')}+\Omega^{j| (p'+1)}=0\,.
\end{align}
The procedure can be iterated again and again, till
no more cocycles can be found. In this way, we obtain the largest FDA associated with the initial Lie algebra.

\medskip

\noindent \underline{{\bf Extension to supersymmetric theories}}: In the supersymmetric case, a set of nontrivial cocycles is generally present in superspace due to the existence of Fierz identities obeyed by the wedge products of gravitino $1$-forms. \\
The $1$-form fields one starts from are the vielbein $V^a$, the gravitino $\psi$, the spin connection $\omega^{ab}$ and, possibly, a set of gauge fields.
We request that the FDA is described in terms of fields living on \emph{ordinary superspace}, whose cotangent space is spanned by the supervielbein only. This corresponds to the physical request of a principal superbundle structure, with superspace as base space, the rest of the fields spanning the fiber. 
This fact implies horizontality of the FDA, corresponding to gauge invariance: All the fields but the supervielbein must be excluded from the construction of the cochains. This corresponds to require the CE-cohomology to be restricted to the $H$-relative CE-cohomology.

\medskip

For instance, the FDA underlying $D=11$ supergravity \cite{DAuria:1982uck,Castellani:1991eu} is
\begin{align}\label{d11fda}
    \mathcal{R}^{ab} & := d \omega^{ab} + {\omega^a}_c \wedge \omega^{cb} = 0 \,, \\
    \mathcal{R}^a & := D V^a - \frac{i}{2}\bar \Psi \wedge \Gamma^a \Psi = 0 \,, \\
    \rho & := D \Psi = 0 \,, \\
    F^{(4)} & := dA^{(3)}-\frac{1}{2} \bar\Psi \wedge \Gamma_{ab}\Psi \wedge V^a \wedge V^b = 0 \,, \\
    F^{(7)} & := dB^{(6)} - 15 A^{(3)} \wedge d A^{(3)} - \frac{i}{2} \bar\Psi \wedge \Gamma_{a_1\ldots a_5} \Psi \wedge V^{a_1} \wedge \cdots \wedge V^{a_5} = 0 \,,
\end{align}
where $\Psi$ is a 32-components Majorana spinor, the $\Gamma$'s are Dirac gamma matrices in eleven dimensions, and $a,b,\ldots\in \{0,1,\ldots,10\}$. On the l.h.s. of \eqref{d11fda} we have defined the super field strengths, whose vanishing defines the vacuum. The latter corresponds to the r.h.s. of \eqref{d11fda}, that is the FDA. As we can see, the fully extended FDA \eqref{d11fda} includes also a (``magnetic'') 6-form gauge potential $B^{(6)}$, related to the Hodge-dual of the field strength of $A^{(3)}$ on spacetime. The $d^2$-closure of this FDA is a consequence of 3-gravitinos Fierz identities in $D = 11$. 

Let us conclude by mentioning that it was shown in \cite{DAuria:1982uck} that the FDA can be traded for an ordinary Lie superalgebra (written in its dual Maurer-Cartan formulation), namely in terms of 1-form gauge fields valued in nontrivial tensor representations of
Lorentz group $S\!O(1,10)$, allowing the disclosure of the so-called \emph{hidden superalgebra} underlying the FDA \eqref{d11fda}. This superalgebra, at least locally, describes a supergroup manifold (\emph{hidden supergroup}, which is an ordinary Lie supergroup, of the FDA) which could be considered at the group-theoretical starting point for a construction of the supergravity theory. In this case, we say that the FDA has been trivialized in terms of hidden 1-form fields (always referring, actually, to its dual description). 

The procedure was done explicitly, decomposing the 3-form $A^{(3)}$ in terms of a set of trilinear (wedge) products of 1-forms and requiring the decomposition to fulfill the integrability of the original FDA equation for $A^{(3)}$, that is $d(A^{(3)}_{\text{dec.}})-\frac{1}{2} \bar\Psi \wedge \Gamma_{ab}\Psi \wedge V^a \wedge V^b = 0$. This prescription, in principle, can be applied to any FDA. However, there is no strong reason to think that the trivialization of a FDA, meaning the hidden superalgebra obtained, is unique. 
Case by case analysis has shown, for now, that among the new 1-forms needed to ensure that the given decomposition reproduces the integrable equation of a FDA, there appears (at least) one extra spinor 1-form field. 
The role of these spinor 1-forms, investigated in \cite{Andrianopoli:2016osu,Andrianopoli:2017itj,Ravera:2021sly}, is to ensure that the new 1-form fields introduced to trivialize the FDA do not carry physical degrees of freedom (the dependence on the new coordinates introduced is completely factorized and the curvatures are horizontal). It would be interesting to understand these results within a conceptually clearer  mathematical framework. 

\section{Conclusion}
\label{Conclusion}

We have here argued for the foundational relevance of Cartan geometry and Cartan supergeometry for gauge field theories of gravity and supergravity. 
Actually, this is true for the class of theories where the gauge potential and superpotential are 1-forms and symmetries are 0-forms. The class of theories with higher form fields and symmetries, such as encountered  supergravity (as just mentioned) and string models, should be understood via higher geometry. 
In that context, a popular approach relies on the (1-)categorification of the notion of connection and its extension to $n$-categories ($n \geq 2$) \cite{Baez-Huerta2011, jurvco2015semistrict,jurvco2016higher}. 
A related algebraic approach, corresponding to a trivial geometric structures, relies on homotopy ($L_\infty$-)algebras \cite{Kim_Seamann2019}.  
Quite naturally, higher Cartan (super)geometry ought to be the relevant mathematical foundations for this broader class. 
It is still a largely underdeveloped area of enquiry, and contributions to it promise much clarifying insight into supergravity and string theory -- see e.g. the recent \cite{Eder-Huerta-Noja2023}. 
From a mathematical perspective, higher Cartan geometry may be considered a higher extension of 
 \emph{natural geometric structures} as defined in \citep{Kolar-Michor-Slovak}, and therefore another chapter in the study of the geometry of  (super)manifolds. 
  In~particular, its  flat limit, higher Klein geometry, would lay the basis for a ``\emph{higher Erlangen program}". 

Supergeometry is the mildest form of non-commutativity. One may submit that non-commutative Cartan geometry would be a beautiful subject to investigate. Here again, several approach to non-commutative (NC) geometry exist: à la Connes via spectral triples \cite{Connes-Marcolli, Connes2010}, à la Dubois-Violette via derivations \cite{Dubois-Violette_kerner_Madore1990a, Dubois-Violette-Kerner-Madore1990b}, via Hopf algebras (a.k.a. ``quantum groups") and deformations etc., giving as many possible incarnations of NC Cartan geometry.
In the latter case, it would generalise the NC Klein geometry of spaces otherwise known as quantum homogeneous spaces
\cite{Kasprzak:2012aa, Schneider1990, OBUACHALLA2016}, and would also give an encompassing framework for NC gauge field theory on $\kappa$-Minkowski space, the homogeneous space for the $\kappa$-Poincaré group \cite{Wallet-et-al}. More generally, NC Cartan geometry is a natural mathematical arena for  models of quantum gravity (at least ``phenomenological" ones). A topic that has been, and remains, a driving motivation at the  frontier of knowledge in theoretical and mathematical physics. 

\medskip

\section*{Acknowledgment}

L.R. would
like to thank the DISAT of the Polytechnic of Turin and the INFN for financial support and L. Andrianopoli for stimulating discussions. \\
J.F. is supported by the OP J.A.C. MSCA grant, number CZ.02.01.01/00/22\_010/0003229, co-funded by the Czech government Ministry of Education, Youth \& Sports and the EU,
as well as  the Austrian Science Fund (FWF), [P 36542].

{
\small
 \bibliography{superbib}
}

\end{document}